\documentclass[12pt,noshowpacs,nofootinbib,notitlepage,amsmath,superscriptaddress]{revtex4-1}
\usepackage{setspace}
\usepackage{hyperref}
\usepackage{amsfonts}
\usepackage{appendix}
\usepackage{mathtools}
\usepackage{ulem}
\usepackage{ifthen}
\newboolean{PrintVersion}
\setboolean{PrintVersion}{false}
\usepackage{amsmath,amssymb,amstext} 
\usepackage[pdftex]{graphicx} 
\usepackage[caption=false]{subfig}
\usepackage{wrapfig}
\usepackage{cleveref} 
\usepackage{enumitem}
\usepackage{float}
\usepackage{mathtools}
\usepackage{xcolor}
\usepackage[export]{adjustbox}
\usepackage{float}
\usepackage{booktabs}

\ifthenelse{\boolean{PrintVersion}}{   
\hypersetup{	
    citecolor=black,%
    filecolor=black,%
    linkcolor=black,%
    urlcolor=black}
}{} 

\let\origdoublepage\cleardoublepage
\newcommand{\clearemptydoublepage}{%
  \clearpage{\pagestyle{empty}\origdoublepage}}
\let\cleardoublepage\clearemptydoublepage

\begin{document}

\title{Gravitational Wormholes}

\author{Mengqi Lu}
\email{mengqi.lu@uwaterloo.ca}
\affiliation{Department of Physics and Astronomy, University of Waterloo, Waterloo, Ontario, Canada, N2L 3G1}

\author{Jiayue Yang}
\email{j43yang@uwaterloo.ca}
\affiliation{Department of Physics and Astronomy, University of Waterloo, Waterloo, Ontario, Canada, N2L 3G1}
\affiliation{Perimeter Institute for Theoretical Physics, 31 Caroline St. N., Waterloo, ON N2L 2Y5, Canada}

 \author{Robert B. Mann}
\email{rbmann@uwaterloo.ca}
\affiliation{Department of Physics and Astronomy, University of Waterloo, Waterloo, Ontario, Canada, N2L 3G1}

\begin{abstract}{Spacetime wormholes are evidently an essential component of the  construction of a time machine. Within the context of general relativity, such objects require, for their formation, exotic matter---matter that violates at least one of the standard energy conditions. Here, we explore the possibility that higher-curvature gravity theories might permit the construction of a wormhole without any matter at all.   In particular, we consider the simplest form of a generalized quasi topological theory in four spacetime dimensions, known as Einsteinian Cubic Gravity.  This theory has a number of promising features that make it an interesting phenomenological competitor to general relativity, including having non-hairy generalizations of the Schwarzschild black hole and linearized equations of  second order around maximally symmetric backgrounds. By matching series solutions near the horizon and at large distances, we find  evidence that strong asymptotically AdS wormhole solutions can be constructed, with strong curvature effects ensuring that the wormhole throat  can exist.   }
\end{abstract}

\maketitle

\newpage

\section{Introduction}

It has long been known that if spacetime is to have closed timelike curves in some local regions \cite{Morris:1988tu}, then wormholes are an essential part of this construction \cite{Morris:1988cz,Visser:1995cc}. However, a key characteristic of such objects is that they require exotic matter that does not respect the energy conditions.  Despite the challenges presented in constructing wormholes \cite{Visser:1992tx}, the~search nevertheless continues in the hopes of evading the constraints imposed by quantum physics in Einsteinian geometries \cite{Ford:1995wg}.

Much effort has gone into exploring modified gravity to this end \cite{Harko:2013yb,Nandi:1997mx,Yue:2011cq,Lobo:2010sb,Sushkov:2011zh,Lobo:2009ip,
Garcia:2010xb,MontelongoGarcia:2010xd,Anabalon:2019lzc}.  The higher-curvature Lovelock theories \cite{Lovelock1,Lovelock2} have been of particular interest
\cite{Bhawal:1992sz,Wang:1996xn,Shang:1999xs,Maeda:2008nz,Dehghani:2009zza,Mehdizadeh:2016nna,Mehdizadeh:2012zz,Mehdizadeh:2015jra}
since the field equations are of second order in the metric components.  However, non-trivial solutions exist only in spacetime dimensions $D \geq 5$. Curiously, quasi-topological 
gravities \cite{Quasi2,Quasi,Oliva:2011xu,Oliva:2012zs,Dehghani:2011vu,Cisterna:2017umf} could also be considered for wormhole solutions, but none have been obtained to date.  However, such objects would also exist only in $D \geq 5$ dimensions.

Both Lovelock and quasi-topological theories 
have been shown to be   particular cases of a more general class called   Generalized Quasi-Topological  gravity (GQTG) \cite{Hennigar:2017ego,PabloPablo3,Ahmed:2017jod,PabloPablo4,Bueno:2019ycr}.  These theories  are characterized by 
second-order linearized equations around maximally symmetric backgrounds and 
admit  single-function ($g_{tt}g_{rr}=-1$) non-hairy generalizations of the Schwarzschild black hole. 
These theories are ghost-free on constant-curvature backgrounds, but, on a generic background, will have ghosts.  However, such ghosts cannot escape to infinity in spacetimes that are asymptotically of constant curvature. 
{{The effects of the additional} 
 degrees of freedom in GQTGs
have not been fully explored, but it is known that 
they significantly modify the thermodynamics of black holes
for small masses \cite{PabloPablo4} and,  in the cubic and quartic cases, exhibit an number of interesting features}
\cite{PabloPablo4,Hennigar:2017umz,Mir:2019rik,Mir:2019ecg,Li:2017ncu,Li:2017txk,Li:2018drw}. A~comprehensive list of their properties has been given
\cite{Bueno:2019ycr,Bueno:2019ltp}.
A key advantage of GQTGs is that they have non-trivial field equations (and solutions) in $D=4$ dimensions.

In this paper, we carry out the first investigations of wormhole solutions in $D=4$ Generalized Quasi-Topological  gravity.  For specificity, we shall consider the simplest GQTG,   a theory known as Einsteinian Cubic Gravity (ECG) \cite{Hennigar:2017ego, Bueno:2016xff}, whose action is of the form  
\begin{equation}\label{action}
    \mathcal{I}=\dfrac{1}{16\pi}\int d^4x \sqrt{-g}(-2\Lambda_0+ R+\alpha \mathcal{P}+\beta \mathcal{C}+\gamma \mathcal{C}'),
\end{equation}
with $\alpha$, $\beta$ and $\gamma$ being coupling constants, and three densities that are cubic in the Riemann curvature, given by 
 \begin{equation}
    \mathcal{P}=12R_{a\ b}^{\ c\ d}R_{c\ d}^{\ e\ f }R_{e\ f}^{\ a\ b}+R_{ab}^{\ \ cd}R_{cd}^{\ \ ef}R_{ef}^{\ \ ab}-12R_{abcd}R^{ac}R^{bd}+8{R^b}_{a}{R^c}_{b}{R^a}_{c},
\end{equation}
\begin{equation}
    \mathcal{C}=\dfrac{1}{2}R{R_a}^{b}{R_b}^{a}-2R^{ac}R^{bd}R_{abcd}-\dfrac{1}{4}RR_{abcd}R^{abcd}+R^{de}R_{abcd}R^{abc}_{\ \ \ \  e},
\end{equation}
\begin{equation}
    \mathcal{C}'={R_a}^b{R_b}^c{R_c}^a-\dfrac{3}{4}R{R_a}^b{R_b}^a+\dfrac{1}{8}R^3.
\end{equation}
 {For} 
 a general static spherically symmetric ansatz (GSSS)
 \begin{equation}\label{GSSS}
    ds^2=-f(r)dt ^2+\dfrac{1}{N(r)f(r)}dr^2+r^2(d\theta^2+\sin^2{\theta}d\phi^2),   
\end{equation}
the densities $\mathcal{C}$ and $\mathcal{C'}$ 
do not contribute in a linearly independent way to the field equations.~Both of these terms  become trivial when the metric function $N(r)$ is a constant. This~situation is the one generally considered in ECG and clearly does not admit  a vacuum wormhole.~{Note that,  regardless of the values that the higher-curvature couplings take, 
the~Einstein--AdS limit of the theory at large distances is preserved 
(albeit with a modified cosmological constant),
since the contributions from the cubic terms fall off much more rapidly than the contributions from the Einstein--Hilbert part of the action if AdS asymptotic behaviour is imposed as a requirement.  }

We therefore  consider the GSSS ansatz with $N'\neq 0$.~We find that this situation is possible  if the spacetime has a spherical deficit/surfeit angle in the  asymptotic large-$r$ region. We shall specifically consider solutions whose metric functions have the asymptotic~form
\begin{equation}\label{bc1}
    g_{tt}\sim -\bigg(\dfrac{r^2}{l^2}+1+\delta\bigg)+\mathcal{O}(r^{-1}), \quad g^{rr}\sim \dfrac{r^2}{l^2}+1+\delta+\mathcal{O}(r^{-1})
\end{equation}
where $l$ is the AdS length scale and $\delta\neq0$   parametrizes the 
deficit/surfeit angle.  

 We find that Einsteinian Cubic Gravity---and, by implication, higher-order GQTGs---admits wormhole solutions that are purely gravitational without any exotic matter.  The solutions that we obtain are
asymptotically anti-de Sitter, with a spherical deficit angle resembling that of a global monopole.  Unlike other solutions with radial symmetry, these solutions have non-zero values for the coupling parameter $\beta$.   We find that including $\beta$ provides a sufficiently large number of parameters to match the series solutions for the two metric functions over a broad range of radii at which the matching takes place.

\section{ The Non-Linear ODE System }\label{sec:4dECG}

For the ansatz \eqref{GSSS}, we find that the two independent field {equations are}
\begin{equation}
\begin{array}{ll}
&-\dfrac{1}{8r^5} \Bigg\{ 8r \Big( r^2 (-1 + r^2 \Lambda_0) + 6\alpha N^2 f'^2 (-1 + r^2 f' N') + r N f' (r^2 + 12\alpha f' N') \Big) 
\vspace{1.5ex}\\
&+ 2f \Big[ 4r^2 N' (r^2 + 15\alpha f' N') + r N \Big( 3r^2 (74\alpha + 5\beta) f'^2 N'^2 + 4r ( r + 48\alpha N' f'' ) \vspace{1.5ex} \\
&- 24f' \Big( (8\alpha - \beta) N' - 8r\alpha N'' \Big) \Big) + 24r\alpha N^3 \Big( 4f'^2 + r^2 f''^2 + r f' ( -4f'' + r f''' ) \Big) \vspace{1.5ex}\\
&+ 12N^2 \left( f' \Big( 8\alpha + r^3 (22\alpha + \beta) N' f'' \Big) + 8r^2\alpha f'^2 \Big( -3N' + r N'' \Big) + 4r\alpha \Big( -2f'' + r f''' \Big) \right)\Big]\vspace{1.5ex} \\
&+ 6f^3 \Big[ r^2 N'^2 \Big( -4\alpha N' + r (8\alpha + \beta) N'' \Big) + 4N^2 \Big( 2\beta N' - r \Big( (-4\alpha + \beta) N'' + 4r\alpha N''' \Big) \Big)\vspace{1.5ex}\\
&+ r N \Big( (8\alpha - 7\beta) N'^2 + 2r^2 (8\alpha + \beta) N''^2 + 2r N' \Big( -20\alpha N'' + r (8\alpha + \beta) N''' \Big) \Big) \Big] \vspace{1.5ex}\\
&+ 3f^2 \biggl[ r N' \Big( 6\beta N' + r^2 (28\alpha + 3\beta) f' N'^2 + 16r\alpha N'' \Big) - 32\alpha N^3 \Big( 2f' + r ( -2f'' + r f''' ) \Big)\vspace{1.5ex}\\
&+ 2N \Big( 2r^2 N'^2 \Big( -52\alpha f' + r (26\alpha + 3\beta) f'' \Big) + N' \Big( -8\beta + r^3 (108\alpha + 11\beta) f' N'' \Big)\vspace{1.5ex}\\
&+ 4r \Big( (-4\alpha + \beta) N'' + 4r\alpha N''' \Big) \Big) + 4r N^2 \Big[ r \Big( r (12\alpha + \beta) f'' N'' + N' \Big( -48\alpha f'' + r (8\alpha + \beta) f''' \Big) \Big)\vspace{1.5ex}  \\
&+f' \Big(6(8\alpha - \beta) N' + 4r\alpha ( -12N'' + r N''' ) \Big) \Big] \biggr] \Bigg\}=0
\label{FE1}
\end{array}
\end{equation}
and
\begin{equation}
\begin{array}{ll}
&\dfrac{1}{8r^5} \Biggl\{ -8r \Big( r^2 (-1 + r^2 \Lambda_0) + 6\alpha N^2 f'^2 (-1 + r^2 f' N') + r N f' (r^2 + 12\alpha f' N') \Big)\vspace{1.5ex} \\
&+ 6\beta f^3 N \Big( r N'^2 + N ( -8N' + 4r N'')\Big) - 2f \Bigg[ -12r^2\alpha f' N'^2 + r N \bigg( 3r^2 (26\alpha + 3\beta) f'^2 N'^2 \vspace{1.5ex}\\
&+ 4r \Big( r + 12\alpha N' f'' \Big) + 24f' \Big( (-2\alpha + \beta) N' + 2r\alpha N'' \Big) \bigg) + 24r\alpha N^3 \Big( 4f'^2 + r^2 f''^2  \vspace{1.5ex} \\
&+ r f' ( -4f'' + r f''' ) \Big) + 12N^2 \bigg( f' \Big( 8\alpha + r^3 (12\alpha + \beta) N' f'' \Big) + 2r^2\alpha f'^2 ( -8N' + r N'' )  \vspace{1.5ex}\\
&+ 4r\alpha ( -2f'' + r f''' ) \bigg) \Bigg] + 3f^2\Bigg[  r N'^2 \Big( 2\beta + r^2 (8\alpha + \beta) f' N' \Big) - 2N \Bigg( 2r^2 N'^2 \Big( -4\alpha f' \vspace{1.5ex} \\
&+ r (8\alpha + \beta) f'' \Big) + 4r\beta N'' + N' \Big( -8\beta + 3r^3 (8\alpha + \beta) f' N'' \Big) \Bigg) + 32\alpha N^3 \Bigg( 2f' + r  \vspace{1.5ex}\\
&\Big( -2f'' + r f''' \Big) \Bigg) - 4rN^2 \Bigg( 2f' \Big( (8\alpha - 3\beta) N' - 8r\alpha N'' \Big) + r \Big( r (8\alpha + \beta) f'' N'' + N'  \vspace{1.5ex} \\
&\Big( -16\alpha f'' + r (8\alpha + \beta) f''' \Big) \Big) \Bigg)\Bigg] \Biggr\}=0
\label{FE2}
\end{array}
\end{equation}
where, without loss of generality, we can set $\gamma=0$ in \eqref{action}, since its inclusion simply reproduces the preceding equations but with $\beta \to \beta + \gamma/2$.

If $N(r)$ is constant, then these equations become  linearly dependent, and a wormhole solution is not possible since there will be a single metric function whose largest root  corresponds to the horizon of a black hole.

We can see this by considering 
the series expansions 
\begin{equation}\label{anz1}
    f_{\infty}=1-\dfrac{\Lambda }{3}r^2+\sum_{i=1}\dfrac{\Tilde{f}_i}{r^i}, \quad N_{\infty}=1+\sum_{i=1}\dfrac{\Tilde{N}_i}{r^i}
\end{equation}
in the asymptotically distant region at large $r$, 
where asymptotically flat solutions have $\Lambda=0$.  Inserting these into the field Equations \eqref{FE1} 
and \eqref{FE2}
yields

\begin{align}
h(\Lambda) &= 0
\quad \Tilde{N_i}=0   \label{hNeqn}  \\
   f_{\infty} &=1-\dfrac{\Lambda}{3} r^2-\dfrac{r_h}{r}+\dfrac{(54-28\Lambda r^2)\alpha r_h^2}{h'(\Lambda)r^6}-\dfrac{138\alpha r_h^3+20192\Lambda^2 \alpha^2 r_h^3}{3h'^2r^{7}}+\mathcal{O}(r^{-9})
 \label{finfeqn}  
\end{align}
in the limit   $r\rightarrow \infty$, where 
\begin{equation}\label{heqn}
    h(x)\equiv\frac{8\alpha}{9} x^3+x-\Lambda_0
\end{equation}
and
 \begin{equation}\label{hprm}
    \Lambda=-\dfrac{3}{l^2}\equiv-3L,  \quad h'(\Lambda) = \dfrac{d h}{dx}\bigg|_{\Lambda}
    =\dfrac{8\alpha}{3}\Lambda^2+1 =24\alpha L^2+1
\end{equation}
and all $\beta$-dependent terms vanish.

From these formulae, we observe several aspects.  First, a power-series solution
in $1/r$ implies only a single independent function $f(r)$, which is the hallmark of GQTGs.  Second,
for asymptotically flat solutions $\Lambda=0$, which in turn implies $\Lambda_0 = 0$, we also have $h' = 1$, and so the asymptotically flat solution can be immediately obtained from \eqref{finfeqn} by setting $\Lambda=0$. However, note that the converse is not true: even if $\Lambda_0 = 0$, it is possible to have asymptotically de Sitter solutions with $\Lambda = 3/\sqrt{8|\alpha|}$ provided that $\alpha<0$.

We seek solutions that have the asymptotic form \eqref{bc1},  where $N(r)$ is not constant so as to obtain wormhole solutions. 
The presence of the wormhole needs to be manifest at large-$r$ in a way that differs from that of a spherically symmetric star or black hole.   
To this end, we consider the ansatz
\begin{equation}\label{anz2}
f_{\infty}=K+Lr^2+\sum_{i=1}\dfrac{a_i}{r^i}, \quad N_{\infty}=1+\sum_{i=1}\dfrac{b_i}{r^i}
\end{equation}
where the quantity $K=1+\delta$ parameterizes
a spherical deficit/surfeit angle produced by the wormhole.  The effect is analogous to that produced by a global monopole \cite{Barriola:1989hx,Shi:1991yto}. Far~from the wormhole, all light rays are deflected by the same angle regardless of their impact parameter.

Inserting the ansatz \eqref{anz2} into the field equations 
yields
\begin{align}\label{balpcond}
    &b_1 L (24L^2\alpha + 1) = 0 
\end{align}
from both equations to leading order. This equation is satisfied by choosing either $b_1=0$ or     $(24 L^2\alpha + 1)=0$.  However, it is straightforward to show that the next order forces $b_1=0$ regardless of the value of $(24 L^2\alpha + 1)$.  However, if this latter quantity is non-zero, then it is straightforward to show that there is no deficit angle, and all $b_i$ coefficients must vanish as per the discussion above.

Setting $(24 L^2\alpha + 1)=0$ then yields
 the non-{trivial solution}
\begin{equation}
\begin{array}{ll}
    N_{\infty}&= 1-\dfrac{5 a_1 }{2 L r^3 \left(9 \beta  L^2-2\right)}+\dfrac{135 a_1^2 \left(\beta  L^2 \left(99 \beta  L^2-49\right)+6\right)}{16 (K-1) L r^4 \left(9 \beta  L^2-2\right)^3} \vspace{1.5ex}\\
        &+\dfrac{a_1 \left(192 (K-1)^2 K \left(2-9 \beta  L^2\right)^2-27 a_1^2 L \left(3 \beta  L^2 \left(3972 \beta  L^2-2207\right)+919\right)\right)}{64 (K-1)^2 L^2 r^5 \left(9 \beta  L^2-2\right)^3}+\cdots      
   \end{array}    
\end{equation}
\begin{equation}
\begin{array}{ll}
    f_{\infty} &=    K+L r^2+\dfrac{a_1}{r}-\dfrac{9 a_1^2 \left(828 \beta ^2 L^4-363 \beta  L^2+37\right)}{32 (K-1) r^2 \left(9 \beta  L^2-2\right)^2}\vspace{1.5ex}\\
    &+\dfrac{a_1 \left(9 a_1^2 L \left(3 \beta  L^2 \left(9 \beta  L^2 \left(9932 \beta  L^2-6829\right)+13340\right)-2641\right)\right)}{224 (K-1)^2 L r^3 \left(9 \beta  L^2-2\right)^3}\vspace{1.5ex}\\
    &+\dfrac{a_1 \left(-32 (K-1)^2 K \left(2-9 \beta  L^2\right)^2 \left(9 \beta  L^2-6\right)\right)}{224 (K-1)^2 L r^3 \left(9 \beta  L^2-2\right)^3}+\cdots 
\end{array}
\end{equation}
where $\Lambda_0=-2L$ from \eqref{hNeqn} and \eqref{heqn}.

We now see that the condition 
$h'(\Lambda) =0$ from \eqref{hprm} allows for $N$ to be a  non-constant function, opening up the possibility of obtaining a wormhole solution.
We pursue this in the next section.

\section{Series Solutions}

Anticipating the asymptotic behaviour \eqref{anz2},  we rewrite the general static spherically symmetric (GSSS) ansatz 
\eqref{GSSS} in the form 
\begin{equation}\label{GSSSx}
    ds^2=-\dfrac{r_0^2 g(x)}{(1-x)^2}d\Tilde{t}^2+\dfrac{r_0^2 dx^2}{n(x)g(x)(1-x)^2} +\dfrac{r_0^2 }{(1-x)^2}(d\theta^2+\sin^2{\theta}d\phi^2),
\end{equation}
using the coordinate transformation 
$$
x=1-r_0/r \qquad \Tilde{t}=t/r_0
$$
where the metric functions $n(x)$ and $g(x)$, defined on $x \in [0,1]$, are
\begin{align}
    n = N \quad  g = \dfrac{ f  {r_0^2}}{r^2}
\end{align}
with $r_0$ a positive constant.

For    wormhole solutions   \cite{Morris:1988tu}, these continuous functions   must be everywhere positive in the interior of the domain, with $n$ vanishing and $g$ having a finite positive value  at $x=0$, which   locates the position of the wormhole throat. Under this map, $r\to \infty$ is compactified to $x=1$.  With this   new ansatz, the boundary condition \eqref{bc1} is equivalent to   
\begin{equation}\label{asypansatz}
    g\sim \dfrac{r_0^2}{ l^2}+\dfrac{(1+\delta)r_0^2}{r^2}+\mathcal{O}(r^{-3}), \quad n\sim 1+\mathcal{O}(r^{-3})
\end{equation}
as $x$ approaches $1$. 
The effect of $\delta$ is analogous to that produced by a global monopole~\cite{Barriola:1989hx,Shi:1991yto}, which  deflects light rays by the same angle regardless of their impact parameter.

The advantage of the ansatz \eqref{GSSSx} is clear---it   compactifies the domain so that 
numerical and semi-analytic solutions 
  become more easily attainable. 
We now employ this  ansatz  to obtain series solutions for the functions $n$ and $g$. The field equations for 
$g(x)$ and $n(x)$ are given in Appendix~\ref{AppA}.

\subsection{Large-\texorpdfstring{$r$}\ \  Solution} 

To obtain solutions asymptotic to \eqref{asypansatz},   we substitute the formal series 
\begin{equation}
    g=a_0+\sum_{n=2}^{\infty}a_n (1-x)^n, \quad n=1+\sum_{n=3}^{\infty}b_n (1-x)^n, \quad a_0=r_0^2/l^2\equiv L r_0^2, \quad a_2=1+\delta
\end{equation}
 into the equations and solve them order by order
in $(1-x)$.~Note that we have set $b_1=b_2=0$ due to the discussion
following condition \eqref{balpcond}. 
The lowest two orders yield two constraints $h(a_0)=0$ and $(24\alpha L^2+1)(a_2-1)=0$.  The first of these simply defines $\Lambda_0$ in terms of the other parameters.  
Solutions with a vanishing deficit $a_2=1$
satisfy the second constraint but force 
$ {b_{n\geq 3}}=0$ or, in other words, $n=1$.
The only alternative non-trivial solution occurs when $a_2\neq 1$, {yielding }

\begin{equation}
\setlength{\arraycolsep}{5pt}
\begin{array}{ll}
g=&L r_0^2+a_2 (1-x)^2-\dfrac{2}{5} b_3 L r_0^2 (1-x)^3 \left(9 \beta L^2-2\right)\vspace{1.5ex}\\
&-\dfrac{9 b_3^2 L^2 r_0^4 (1-x)^4 \left(828 \beta ^2 L^4-363 \beta L^2+37\right)}{200 (a_2-1)}\vspace{1.5ex}\\
&+
\dfrac{3 b_3 (1-x)^5}{3500 (a_2-1)^2}
\Biggl\{
\left(200 a_2^3 \left(3 \beta L^2-2\right)-400 a_2^2 \left(3 \beta L^2-2\right)+200 a_2 \left(3 \beta L^2-2\right)\right)\vspace{1.5ex}\\
&
\qquad + \left(3 b_3^2 L^3 r_0^6 \left(-268164 \beta ^3 L^6+184383 \beta ^2 L^4-40020 \beta L^2+2641\right)\right) \Biggr\}\vspace{1.5ex}\\
&-\dfrac{b_3^2 L r_0^2 (1-x)^6 }{4480000 (a_2-1)^3}\Bigg\{\left(-43200 (a_2-1)^2 (1803 a_2+85) \beta ^2 L^4\right.\vspace{1.5ex}\\
&+13200 (a_2-1)^2 (3063 a_2+545) \beta L^2-400 (a_2-1)^2 (13371 a_2+3535)\vspace{1.5ex}\\
&+72626980752 \beta ^4 b_3^2 L^{11} r_0^6-68916687624 \beta ^3 b_3^2 L^9 r_0^6+23598134433 \beta ^2 b_3^2 L^7 r_0^6\vspace{1ex}\\
&-3385644966 \beta b_3^2 L^5 r_0^6+164339937 b_3^2 L^3 r_0^6\left.\right)\Biggr\}\vspace{1.5ex}\\
&+\dfrac{b_3 (1-x)^7 }{392000000 (a_2-1)^4 L r_0^2}\Biggl\{36000 a_2 b_3^2 L^3 r_0^6 \left(5699484 \beta ^3 L^6-3438357 \beta ^2 L^4\right.\vspace{1.5ex}\\
&\left.+659159 \beta L^2-39732\right) +800 a_2^3 \left(295678404 \beta ^3 b_3^2 L^9 r_0^6-211323051 \beta ^2 b_3^2 L^7 r_0^6\right.\vspace{1.5ex}\\
& \left.+49375923 \beta b_3^2 L^5 r_0^6-3749468 b_3^2 L^3 r_0^6+336000 \beta L^2-448000\right)\vspace{1.5ex}\\
&-800 a_2^2 \left(571755996 \beta ^3 b_3^2 L^9 r_0^6-394347609 \beta ^2 b_3^2 L^7 r_0^6+88894962 \beta b_3^2 L^5 r_0^6\right.\vspace{1.5ex}\\
&\left.-6518172 b_3^2 L^3 r_0^6+84000 \beta L^2-112000\right)-22400000 a_2^6 \left(3 \beta L^2-4\right)\vspace{1.5ex}\\
&+89600000 a_2^5 \left(3 \beta L^2-4\right)-134400000 a_2^4 \left(3 \beta L^2-4\right)\vspace{1.5ex}\\
&+b_3^2 L^3 r_0^6 \left(-170858662247952 \beta ^5 b_3^2 L^{13} r_0^6+207845496209160 \beta ^4 b_3^2 L^{11} r_0^6\right.\vspace{1.5ex}\\
&\left.-98339490752265 \beta ^3 b_3^2 L^9 r_0^6+22351263412410 \beta ^2 b_3^2 L^7 r_0^6-2386812017025 \beta b_3^2 L^5 r_0^6\right.\vspace{1.5ex}\\
&\left.+91235810916 b_3^2 L^3 r_0^6+15680649600 \beta ^3 L^6-22638794400 \beta ^2 L^4+7885507200 \beta L^2\right.\vspace{1.5ex}\\
&\left.-784611200\right)\Biggr\}+\cdots \label{ginf1}
\end{array}
\end{equation}

\begin{equation}
\begin{array}{ll}
n=&1+b_3 (1-x)^3+\dfrac{27 b_3^2 L r_0^2 (1-x)^4 \left(11 \beta L^2-3\right)}{20 (a_2-1)}\vspace{1.5ex}\\
&+\dfrac{3 b_3 (1-x)^5 \left(-400 a_2^3+800 a_2^2-400 a_2+9 b_3^2 L^3 r_0^6 \left(11916 \beta ^2 L^4-6621 \beta L^2+919\right)\right)}{1000 (a_2-1)^2 L r_0^2}\vspace{1.5ex}\\
&+\dfrac{b_3^2 (1-x)^6 }{3500 (a_2-1)^3} \Biggl\{\left(-750 (a_2-1)^2 (105 a_2+16) \beta L^2+125 (a_2-1)^2 (175 a_2+22)\right.\vspace{1.5ex}\\
&+\left.28954908 \beta ^3 b_3^2 L^9 r_0^6-24304212 \beta ^2 b_3^2 L^7 r_0^6+6795981 \beta b_3^2 L^5 r_0^6-633051 b_3^2 L^3 r_0^6\right)\Biggr\}\vspace{1.5ex}\\
&+\dfrac{9 b_3 (1-x)^7 }{6272000 (a_2-1)^4 L^2 r_0^4}\Biggl\{\left.-160 a_2^3 \left(2795796 \beta ^2 b_3^2 L^7 r_0^6-1577271 \beta b_3^2 L^5 r_0^6\right.\right.\vspace{1.5ex}\\
&\left.\left.+222099 b_3^2 L^3 r_0^6+22400\right)+160 a_2^2 \left(5138172 \beta ^2 b_3^2 L^7 r_0^6-2934417 \beta b_3^2 L^5 r_0^6\right.\right.\vspace{1.5ex}\\
&\left.\left.+417978 b_3^2 L^3 r_0^6+5600\right)-480 a_2 b_3^2 L^3 r_0^6 \left(629652 \beta ^2 L^4-379007 \beta L^2+56553\right)\right.\vspace{1.5ex}\\
&\left.+896000 a_2^6-3584000 a_2^5+5376000 a_2^4+3 b_3^2 L^3 r_0^6 \left(54758424816 \beta ^4 b_3^2 L^{11} r_0^6\right.\right.\vspace{1.5ex}\\
&\left.-61487988408 \beta ^3 b_3^2 L^9 r_0^6+25878434823 \beta ^2 b_3^2 L^7 r_0^6-4838200650 \beta b_3^2 L^5 r_0^6\right.\vspace{1.5ex}\\
&\left.+339035463 b_3^2 L^3 r_0^6-24182400 \beta ^2 L^4+11740000 \beta L^2-1398400\right)\Biggr\} +\cdots \label{ninf1}
\end{array}
\end{equation}
 where  
\begin{align}\label{spalp}
 24\alpha L^2+1=0
\end{align} 
and $\Lambda_0$ and $\alpha$ are replaced by expressions in terms of $L$ via the two constraints $h(a_0)=0$  and \eqref{spalp}.

The parameters $a_0 = L r_0^2$, $a_2$ and $b_3$ are   the only free variables in this solution; we~also have  $n(1)=n_0=1$.  Furthermore $\beta$ is an independent coupling parameter.
It can happen in a non-linear system that fewer constants of integration appear in a solution than the differential order of the equations.~Due to the non-linearity, a spontaneous singularity could appear, which means that  the radii of convergence for $g$ and $n$ depend on the initial conditions, namely the values of these parameters.  We do not expect  vanishing radii of convergence for all values of the parameters, since the series with $ {b_3} =  0$ converges to the AdS solution with a deficit.

\subsection{Near-Throat Solution}\label{sec:nearthroatsolution}

There is likewise a near-throat solution for
$(24L^2\alpha + 1) = 0$.~Local solutions near $x=0$ compatible with \eqref{ginf1} and \eqref{ninf1} are necessarily Taylor series. In this case, the desired boundary conditions at the throat require the ansatz to be 
\begin{equation}\label{nthsol}
    n = \sum_{n=1}^{\infty} {A_n}{x^n}
  \qquad  g = B_0 + \sum_{n=1}^{\infty} {B_n}{x^n}, \quad B_0\neq 0,
\end{equation}
which yields two series whose coefficients are fully determined by $A_1$, $B_0$ and $a_0 = L r_0^2$. {We~obtain}

\begin{equation}
\begin{array}{ll}
   n_{th} &=  A_1 x+\dfrac{1}{8 A_1 B_0^2 \left(A_1 B_0 \left(3 \beta  L^2-1\right)-1\right)^2}\vspace{1.5ex} \\
   &\Biggl\{2 A_1^2 B_0^2 \left(1-3 \beta  L^2\right) \left(A_1^2 B_0^2 \left(5-12 \beta  L^2\right)+9 A_1 B_0+4\right)\vspace{1.5ex} \\
  &-8 L^2 {r_0}^4 \left(2 A_1^2 B_0^2 \left(3 \beta  L^2-1\right)+A_1 B_0 \left(7-24 \beta  L^2\right)+12\right) \vspace{1.5ex}\\
  &+48 L^3 {r_0}^6 \left(A_1 B_0 \left(8 \beta  L^2-3\right)-4\right)\Biggr\} x^2 +\cdots \label{nth} 
  \end{array}
\end{equation}
 
\begin{equation}
\begin{array}{ll}
   g_{th} &= B_0  -\dfrac{2 L^2 \left(3 A_1^2 B_0^2 (A_1 B_0+1) \left(\beta -\dfrac{1}{3 L^2}\right)+8 L {r_0}^6+4 {r_0}^4\right)}{A_1^2 B_0 \left(A_1 B_0 \left(3 \beta  L^2-1\right)-1\right)}x +\cdots
   \label{gth}
\end{array}
\end{equation}
{Higher-order} 
 terms are very lengthy and cumbersome to write; we present some of them in Appendix~\ref{AppB}.

Note that in both series solutions 
(near $x=1$ and near $x=0$), the independent parameters 
$r_0$ and $\beta$ always appear as  $L r_0^2$ and  $L^2\beta$. Consequently,  we can set $L=1$ 
 and regard $r_0$ and $\beta$ as independent parameters  without loss of generality.

\section{Matching the Solutions}

As a consequence of the uniqueness of Taylor series, we expect that the Taylor expansions of \eqref{ginf1} and \eqref{ninf1} at $x=0$ 
can be matched with \eqref{nth} and \eqref{gth}
as long as a wormhole solution  (analytic on $[0,1]$) exists for some values of $(\beta, r_0, a_2, b_3, A_1, B_0)$.   We achieve this matching by minimizing the quantity 
 \begin{equation}\label{deltafun}
\Delta(x_0) \equiv
\bigg(\dfrac{\Delta g}{0!}\bigg)^2 + \bigg(\dfrac{\Delta g'}{1!}\bigg)^2 + \bigg(\dfrac{\Delta g''}{2!}\bigg)^2 +\bigg(\dfrac{\Delta n}{0!}\bigg)^2 +\bigg(\dfrac{\Delta n'}{1!}\bigg)^2 +\bigg(\dfrac{\Delta n''}{2!}\bigg)^2 
\end{equation}
 where $x_0$ is the matching point, 
as a function of the parameters $(\beta, r_0, a_2, b_3, A_1, B_0)$, 
where $\Delta F \equiv F_\infty - F_{th}$. 
By matching the second derivatives, we ensure that there are no discontinuities in the Riemann curvature.

\begin{figure}[H]
    \includegraphics[width=12cm]{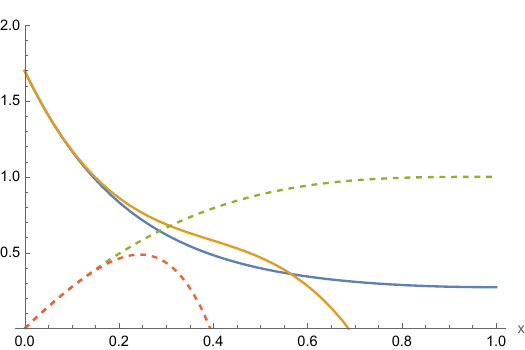}
    \caption{{Plots of the  series solutions $[g(x),n(x)]$  for both large-$r$
    [(solid blue, dashed green]
    and near-throat [(solid orange, dashed red].   The solutions smoothly match with $\Delta=1.03924\times10^{-19}$ at $x~=~x_0~=~0.00505304$ for the parameters  $\beta= 0.505553, a_2 = 0.318022, b_3~=~-0.680002,$ $r_0~=~0.522375, A_1 = 3.03285, B_0 = 1.69888$.}}
    \label{fig:x777}
\end{figure}

\begin{figure}[H]
    \includegraphics[width=12cm]{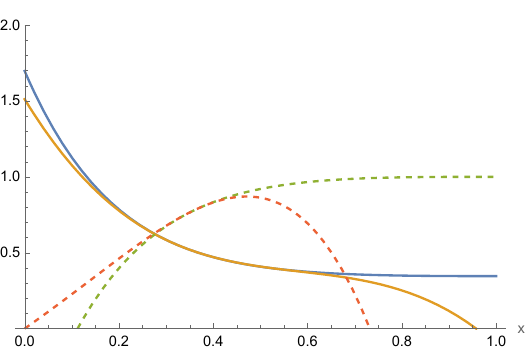}
    \caption{Plots of the  series solutions $[g(x),n(x)]$  for both large-$r$
    [(solid blue, dashed green]
    and near-throat [(solid orange, dashed red].   The solutions
    smoothly match with $\Delta=3.22544\times10^{-15}$ at $x~=~x_0= 0.35156$ for the parameters  $\beta= 0.792546, a_2 = 0.0368723, b_3 = -0.327935,$ $r_0~=~0.588074, A_1 
    =2.1819, B_0 = 1.51153$.}
    \label{fig:x694}
\end{figure}

\begin{figure}[H]
    \includegraphics[width=12cm]{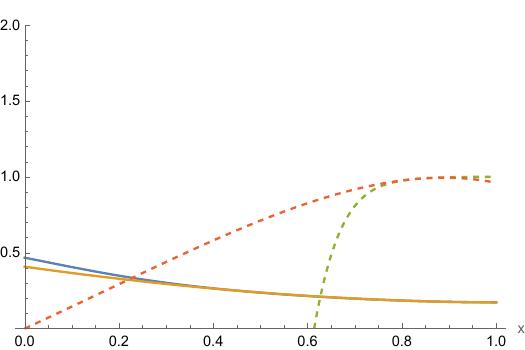}
    \caption{{Plots of the  series solutions $[g(x),n(x)]$  for both large-$r$
    [(solid blue, dashed green]
    and near-throat [(solid orange, dashed red].   The solutions
    smoothly match with $\Delta=3.2155\times10^{-16}$ at $x=x_0=0.83693$ for the parameters  $\beta = 0.0706116, a_2 = 0.34649, b_3 = -3.53033, r_0~=~0.41619,  
A_1~=~1.43351, B_0 = 0.408057$.}}
    \label{fig:x747}
\end{figure}

The presence of the coupling parameter $\beta$, irrelevant for asymptotically AdS solutions (with $K=1$), has a profound effect insofar as it yields a sufficient amount of freedom in the parameter space to minimize $\Delta$ to high precision. The precision of our matching is accurate to one part in $10^{15}$ at worst. Note from
\eqref{spalp} that each solution appears for a specific choice of $\alpha$.
We have found a broad range of wormhole solutions 
using this method. These are illustrated in 
Figures~\ref{fig:x777}--\ref{fig:x747} and, respectively, correspond to matching for small $x_0$, mid-range $x_0$, and large $x_0$.  
\unskip

\section{Conclusions}

We have shown that Einsteinian Cubic Gravity contains wormhole solutions that are purely gravitational.   Unlike wormholes obtained in generic higher-curvature gravity theories, our solutions are in $(3+1)$-dimensions and  require no exotic matter in their~construction.  

In contrast to previous solutions obtained
in the theory, our wormhole solutions require three special characteristics. One is that their asymptotic behaviour is that of AdS spacetime with a global monopole deficit.  The second is that the coupling parameter $\alpha$ is related to the effective cosmological constant via \eqref{spalp}. The third is that the coupling parameter 
$\beta\neq 0$. 
 Our wormhole solutions have no horizons or singularities and so are traversable in principle. 
 Imposing more stringent traversability requirements
 (such as requiring that the gravity at the throat not exceed Earth's gravity) will reduce the range of allowed solutions;  we have not imposed this constraint on the solutions that we have obtained.

Although our series-matching approach has yielded candidate gravitational wormholes, a full wormhole solution to the field Equations \eqref{oseqn1}
and \eqref{oseqn2} remains to be obtained. This can be done numerically, but it presents a computational challenge.~A solution in terms of Tchebyshev polyonomials requires many coefficients to obtain high accuracy.  We were not able to achieve this using the computational resources available.  
  While it is straightforward to apply the shooting method to an ODE system, here, we have a double shooting problem, which is considerably 
more challenging. The solution will necessarily depend on tuning the constants of integration appearing in the local series expansions of the metric functions near $x=1$ such that the integrated solution from $x=1$ satisfies the boundary condition at the other end (or vice versa, beginning at $x=0$). We did attempt such solutions but found that we typically encountered  at least one spontaneous singularity for   $g$ between $x=0$ and $x=1$. Some of these cases might be indicative of a new class of black holes, which merit further investigation.

{Files used to generate the various graphs presented in this paper are available on request.}

\acknowledgments{ 
This work was supported in part by the  Natural Science and Engineering Research Council of Canada. Mengqi Lu  was also supported by the China Scholarship Council. 
We would like to thank  Niayesh Afshordi, Jianhui Qiu, and Robie Hennigar for the helpful discussions.}

\appendix

\section{On-Shell Field Equations}
\label{AppA}

Using the ansatz
\eqref{GSSSx},
field Equations \eqref{FE1} and \eqref{FE2} become, for $\gamma\neq 0$,

\begin{equation}
\begin{array}{ll}\label{oseqn1}
0= & -6 \left(-\left((16 \alpha +2 \beta +\gamma ) n'^2 \left(n'-(x-1) n''\right) (x-1)^3\right)\right.\vspace{1.5ex}\\
&\left.+(16 \alpha +2 \beta +\gamma ) n \left(n'^2-2 (x-1) \left(n''-(x-1) n^{(3)}\right) n'\right.\right.\vspace{1.5ex}\\
&\left.\left.+2 (x-1)^2 n''^2\right) (x-1)^2-64 \alpha n^2 n' (x-1)+64 \alpha n^3\right) g^3\vspace{1ex}\\
&+3 (x-1) \left(-n' \left((56 \alpha +6 \beta +3 \gamma ) g' n'^2+6 (16 \alpha +2 \beta +\gamma ) n'-32 (x-1) \alpha n''\right) (x-1)^3\right.\vspace{1.5ex}\\
&\left.+2 n \left(2 \left(5 (16 \alpha +2 \beta +\gamma ) g'-(x-1) (52 \alpha +6 \beta +3 \gamma ) g''\right) n'^2\right.\right.\vspace{1.5ex}\\
&\left.\left.+\left(-128 \alpha -(x-1) (216 \alpha +11 (2 \beta +\gamma )) g' n''\right) n'\right.\right.\vspace{1.5ex}\\
&\left.\left.-4 (x-1) \left((8 \alpha +2 \beta +\gamma ) n''-8 (x-1) \alpha n^{(3)}\right)\right) (x-1)^2\right.\vspace{1.5ex}\\
&\left.-4 n^2 \left(24 \alpha g'' n'' x^2+2 \beta g'' n'' x^2+\gamma g'' n'' x^2-48 \alpha g'' n'' x-4 \beta g'' n'' x-2 \gamma g'' n'' x\right.\right.\vspace{1.5ex}\\
&\left.\left.-32 \alpha +24 \alpha g'' n''+2 \beta g'' n''+\gamma g'' n''+(x-1) (16 \alpha +2 \beta +\gamma ) n' \left((x-1) g^{(3)}-2 g''\right)\right.\right.\vspace{1.5ex}\\
&\left.\left.+2 g' \left((16 \alpha +2 \beta +\gamma ) n'-(x-1) \left((16 \alpha +2 \beta +\gamma ) n''-4 (x-1) \alpha n^{(3)}\right)\right)\right) (x-1)\right.\vspace{1.5ex}\\
&\left.+128 \alpha n^3 g'\right) g^2\vspace{1ex}\\
&-2 \left(48 \alpha n^3 \left(g''^2+g' g^{(3)}\right) (x-1)^4-12 n^2 \left(2 \left((16 \alpha +2 \beta +\gamma ) n'-8 (x-1) \alpha n''\right) g'^2\right.\right.\vspace{1.5ex}\\
&\left.\left.+\left(-32 \alpha -(x-1) (44 \alpha +2 \beta +\gamma ) n' g''\right) g'+8 (x-1) \alpha \left(2 g''+(x-1) g^{(3)}\right)\right) (x-1)^3\right.\vspace{1.5ex}\\
&\left.-8 n' \left(r_0^4+15 (x-1)^4 \alpha g' n'\right) (x-1)+3 n \left((148 \alpha +5 (2 \beta +\gamma )) g'^2 n'^2 (x-1)^4\right.\right.\vspace{1.5ex}\\
&\left.\left.+8 g' \left((2 \beta +\gamma ) n'-16 (x-1) \alpha n''\right) (x-1)^4+8 \left(r_0^4-16 (x-1)^5 \alpha n' g''\right)\right)\right) g\vspace{1ex}\\
&-16 \left(\left(\Lambda_0 r_0^2-x^2+2 x-1\right) r_0^4+6 (x-1)^4 \alpha n^2 g'^2 \left(g' n'-1\right)\right.\vspace{1.5ex}\\
&\left.-(x-1) n g' \left(r_0^4+12 (x-1)^4 \alpha g' n'\right)\right)
\end{array}
\end{equation} 
and
\begin{equation}
\begin{array}{ll}\label{oseqn2} 
0=&-16 \left(-(x-1) n g' \left(12 \alpha (x-1)^4 g' n'+r_0^4\right)\right.\vspace{1.5ex}\\
&\left.+6 \alpha (x-1)^4 n^2 g'^2 \left(g' n'-1\right)+r_0^4 \left(\Lambda_0 r_0^2-x^2+2 x-1\right)\right)\vspace{1.5ex}\\
&-6 g \left(8 \alpha (x-1)^5 g' n'^2+n \left(8 \left(r_0^4-4 \alpha (x-1)^5 g'' n'\right)+(x-1)^4 (52 \alpha +6 \beta +3 \gamma ) g'^2 n'^2\right.\right.\vspace{1.5ex}\\
&\left.\left.+8 (x-1)^4 g' \left((12 \alpha +2 \beta +\gamma ) n'-4 \alpha (x-1) n''\right)\right)\right.\vspace{1.5ex}\\
&\left.-4 (x-1)^3 n^2 \left(2 g'^2 \left((16 \alpha +2 \beta +\gamma ) n'-2 \alpha (x-1) n''\right)+8 \alpha (x-1) \left((x-1) g^{(3)}+2 g''\right)\right.\right.\vspace{1.5ex}\\
&\left.\left.+g' \left(-32 \alpha -(x-1) (24 \alpha +2 \beta +\gamma ) g'' n'\right)\right)+16 \alpha (x-1)^4 n^3 \left(g''^2+g^{(3)} g'\right)\right)\vspace{1.5ex}\\
&+3 (x-1) g^2 \left((x-1)^3 (16 \alpha +2 \beta +\gamma ) n'^2 \left(g' n'+2\right)+128 \alpha n^3 g'\right.\vspace{1.5ex}\\
&\left.+2 (x-1)^2 n (16 \alpha +2 \beta +\gamma ) \left(g' n' \left(10 n'-3 (x-1) n''\right)-2 (x-1) \left(g'' n'^2+2 n''\right)\right)\right.\vspace{1.5ex}\\
&\left.-4 (x-1) n^2 \left(-32 \alpha +16 \alpha x^2 g'' n''+2 \beta x^2 g'' n''+\gamma x^2 g'' n''-32 \alpha x g'' n''\right.\right.\vspace{1.5ex}\\
&\left.\left.+16 \alpha g'' n''-4 \beta x g'' n''+2 \beta g'' n''-2 \gamma x g'' n''+\gamma g'' n''\right.\right.\vspace{1.5ex}\\
&\left.\left.+2 (16 \alpha +2 \beta +\gamma ) g' \left(n'-(x-1) n''\right)+(x-1) (16 \alpha +2 \beta +\gamma ) \left((x-1) g^{(3)}-2 g''\right) n'\right)\right)\vspace{1.5ex}\\
&-6 g^3 \left((x-1)^3 (16 \alpha +2 \beta +\gamma ) n'^3+3 (x-1)^2 n (16 \alpha +2 \beta +\gamma ) n' \left(n'-2 (x-1) n''\right)+64 \alpha n^3\right)
\end{array}
\end{equation}

\section{Higher-Order Terms of Near-Throat Solutions}
\label{AppB}

Here, we present the first few coefficients of the near-throat solutions 
\eqref{nth} and \eqref{gth} beyond linear order; note that  the present $A_2$ and $B_1$ {appear in the main text.}

\begin{equation}
\begin{array}{ll}
  A_2=&\dfrac{1}{8 A_1 B_0^2 \left(A_1 B_0 \left(3 \beta  L^2-1\right)-1\right)^2}\Biggl\{2 A_1^2 B_0^2 \left(1-3 \beta  L^2\right) \left(A_1^2 B_0^2 \left(5-12 \beta  L^2\right)+9 A_1 B_0+4\right)\vspace{1.5ex} \\
  &-8 L^2 {r_0}^4 \left(2 A_1^2 B_0^2 \left(3 \beta  L^2-1\right)+A_1 B_0 \left(7-24 \beta  L^2\right)+12\right)\\
\end{array}
\end{equation}

\begin{align}
B_1= -\dfrac{2 L^2 \left(3 A_1^2 B_0^2 (A_1 B_0+1) \left(\beta -\dfrac{1}{3 L^2}\right)+8 L {r_0}^6+4 {r_0}^4\right)}{A_1^2 B_0 \left(A_1 B_0 \left(3 \beta  L^2-1\right)-1\right)}
\end{align}

\begin{equation}
\begin{array}{ll}
A_3=& \dfrac{1}{72 A_1^3 B_0^4 \left(\left(3 L^2 \beta -1\right) A_1 B_0-1\right)^5}\Biggl\{221184 L^{12} \beta ^3 A_1^3 B_0^3 {r_0}^{12}\vspace{1.5ex}\\
&+8 L^3 A_1^2 B_0^2 \left(315 A_1^4 B_0^4+1421 A_1^3 B_0^3+2374 A_1^2 B_0^2+1724 A_1 B_0+456\right) {r_0}^6\vspace{1ex}\\
&+6912 L^{11} \beta ^3 A_1^3 B_0^3 \left(32 {r_0}^4+3 \beta A_1^2 B_0^2 (8 A_1 B_0+9)\right) {r_0}^6\vspace{1ex}\\
&+576 L^9 \beta ^2 A_1^2 B_0^2 \left({r_0}^4 \left(5 A_1^2 B_0^2-392 A_1 B_0-448\right)\right.\vspace{1.5ex}\\
&\left.-3 \beta A_1^2 B_0^2 \left(133 A_1^2 B_0^2+257 A_1 B_0+120\right)\right) {r_0}^6\vspace{1ex}\\
&-24 L^7 \beta A_1 B_0 \left(8 {r_0}^4 \left(10 A_1^3 B_0^3-391 A_1^2 B_0^2-936 A_1 B_0-592\right)\right.\vspace{1.5ex}\\
&\left.-3 \beta A_1^2 B_0^2 \left(1675 A_1^3 B_0^3+4642 A_1^2 B_0^2+4056 A_1 B_0+1024\right)\right) {r_0}^6\vspace{1ex}\\
&+8 L^5 \left(8 {r_0}^4 \left(5 A_1^4 B_0^4-127 A_1^3 B_0^3-488 A_1^2 B_0^2-656 A_1 B_0-336\right)\right.\vspace{1.5ex}\\
&\left.-3 \beta A_1^2 B_0^2 \left(1182 A_1^4 B_0^4+4295 A_1^3 B_0^3+5470 A_1^2 B_0^2+2744 A_1 B_0+480\right)\right) {r_0}^6\vspace{1ex}\\
&-3 A_1^4 B_0^4 (A_1 B_0+1)^2 \left(37 A_1^3 B_0^3+98 A_1^2 B_0^2+84 A_1 B_0+24\right)\vspace{1.5ex}\\
&-144 L^{10} \beta ^2 A_1^2 B_0^2 \left(16 (101 A_1 B_0+112) {r_0}^{12}+12 \beta A_1 B_0 \left(3 A_1^2 B_0^2-32\right) {r_0}^8\right.\vspace{1.5ex}\\
&\left.+9 \beta ^2 A_1^3 B_0^3 \left(15 A_1^2 B_0^2-44 A_1 B_0-72\right) {r_0}^4-27 \beta ^3 A_1^5 B_0^5 \left(5 A_1^2 B_0^2-4 A_1 B_0-6\right)\right)\vspace{1.5ex}\\
&+12 L^8 \beta A_1 B_0 \left(64 \left(108 A_1^2 B_0^2+247 A_1 B_0+148\right) \right.\vspace{1.5ex}\\
&\left.{r_0}^{12}+24 \beta A_1 B_0 \left(18 A_1^3 B_0^3+17 A_1^2 B_0^2-190 A_1 B_0-224\right) {r_0}^8\right.\vspace{1.5ex}\\
&\left.+18 \beta ^2 A_1^3 B_0^3 \left(125 A_1^3 B_0^3-301 A_1^2 B_0^2-960 A_1 B_0-480\right) {r_0}^4\right.\vspace{1.5ex}\\
&\left.+27 \beta ^3 A_1^5 B_0^5 \left(-105 A_1^3 B_0^3-30 A_1^2 B_0^2+123 A_1 B_0+32\right)\right)\vspace{1.5ex}\\
&-3 L^6 \left(256 \left(13 A_1^3 B_0^3+46 A_1^2 B_0^2+58 A_1 B_0+28\right) {r_0}^{12}\right.\vspace{1.5ex}\\
&\left.+32 \beta A_1 B_0 \left(18 A_1^4 B_0^4+34 A_1^3 B_0^3-169 A_1^2 B_0^2-442 A_1 B_0-296\right) {r_0}^8\right.\vspace{1.5ex}\\
&\left.+12 \beta ^2 A_1^3 B_0^3 \left(390 A_1^4 B_0^4-741 A_1^3 B_0^3-4112 A_1^2 B_0^2-4016 A_1 B_0-1024\right) {r_0}^4\right.\vspace{1.5ex}\\
&\left.+9 \beta ^3 A_1^5 B_0^5 \left(-891 A_1^4 B_0^4-1245 A_1^3 B_0^3+110 A_1^2 B_0^2+404 A_1 B_0+80\right)\right)\vspace{1.5ex}\\
&+L^2 A_1^2 B_0^2 (A_1 B_0+1) \left(\left(-280 A_1^4 B_0^4+540 A_1^3 B_0^3+4064 A_1^2 B_0^2+5344 A_1 B_0+2016\right) {r_0}^4\right.\vspace{1.5ex}\\
&\left.+3 \beta A_1^2 B_0^2 \left(513 A_1^4 B_0^4+1332 A_1^3 B_0^3+1156 A_1^2 B_0^2+420 A_1 B_0+56\right)\right)\vspace{1.5ex}\\
&+L^4 \left(32 \left(6 A_1^5 B_0^5+17 A_1^4 B_0^4-43 A_1^3 B_0^3-212 A_1^2 B_0^2-308 A_1 B_0-168\right) {r_0}^8\right.\vspace{1.5ex}\\
&\left.+12 \beta A_1^2 B_0^2 \left(270 A_1^5 B_0^5-380 A_1^4 B_0^4-3631 A_1^3 B_0^3-5402 A_1^2 B_0^2-2808 A_1 B_0-480\right) {r_0}^4\right.\vspace{1.5ex}\\
&\left.-9 \beta ^2 A_1^4 B_0^4 \left(953 A_1^5 B_0^5+2390 A_1^4 B_0^4+1823 A_1^3 B_0^3+430 A_1^2 B_0^2+16\right)\right)\Biggr\}
\end{array}
\end{equation}

\begin{equation}
\begin{array}{ll}
B_2&= \dfrac{1}{6 A_1^4 B_0^3 \left(\left(3 L^2 \beta -1\right) A_1 B_0-1\right)^4}\Biggl\{4608 L^{10} \beta ^2 A_1^2 B_0^2 {r_0}^{12}\vspace{1.5ex}\\
& -8 L^3 A_1^2 B_0^2 \left(15 A_1^3 B_0^3+39 A_1^2 B_0^2+34 A_1 B_0+10\right) {r_0}^6 \vspace{1.5ex}\\
&-576 L^9 \beta ^2 A_1^2 B_0^2 \left({r_0}^4 (A_1 B_0-8)-6 \beta A_1^3 B_0^3\right) {r_0}^6 \vspace{1.5ex}\\
&+24 L^7 \beta A_1 B_0 \left(16 {r_0}^4 \left(A_1^2 B_0^2-8 A_1 B_0-12\right)-3 \beta A_1^2 B_0^2 \left(47 A_1^2 B_0^2+42 A_1 B_0+16\right)\right) {r_0}^6 \vspace{1.5ex}\\
&-8 L^5 \left(8 {r_0}^4 \left(A_1^3 B_0^3-8 A_1^2 B_0^2-24 A_1 B_0-12\right)\right.\vspace{1.5ex}\\
&\left.-3 \beta A_1^2 B_0^2 \left(46 A_1^3 B_0^3+81 A_1^2 B_0^2+50 A_1 B_0+8\right)\right) {r_0}^6 \vspace{1.5ex}\\
&+6 A_1^4 B_0^4 (A_1 B_0+1)^4-24 L^8 \beta A_1 B_0 \left(8 (17 A_1 B_0+24) {r_0}^{12}+12 \beta A_1 B_0 (A_1 B_0-4) {r_0}^8\right.\vspace{1.5ex}\\
&\left.+9 \beta ^2 A_1^4 B_0^4 (A_1 B_0-7) {r_0}^4-27 \beta ^3 A_1^5 B_0^5 \left(A_1^2 B_0^2+3 A_1 B_0+1\right)\right) \vspace{1.5ex}\\
&+3 L^6 \left(64 \left(3 A_1^2 B_0^2+8 A_1 B_0+4\right) {r_0}^{12}+16 \beta A_1 B_0 \left(4 A_1^2 B_0^2-15 A_1 B_0-24\right) {r_0}^8\right.\vspace{1.5ex}\\
&\left.+12 \beta ^2 A_1^3 B_0^3 \left(6 A_1^3 B_0^3-33 A_1^2 B_0^2-36 A_1 B_0-16\right) {r_0}^4\right.\vspace{1.5ex}\\
&\left.-27 \beta ^3 A_1^5 B_0^5 \left(10 A_1^3 B_0^3+33 A_1^2 B_0^2+27 A_1 B_0+8\right)\right) \vspace{1.5ex}\\
&-L^2 A_1^2 B_0^2 (A_1 B_0+1) \left(\left(-8 A_1^3 B_0^3+28 A_1^2 B_0^2+56 A_1 B_0+24\right) {r_0}^4\right.\vspace{1.5ex}\\
&\left.+3 \beta A_1^2 B_0^2 \left(26 A_1^3 B_0^3+73 A_1^2 B_0^2+64 A_1 B_0+18\right)\right) \vspace{1.5ex}\\
&+2 L^4 \left(8 \left(-2 A_1^3 B_0^3+7 A_1^2 B_0^2+24 A_1 B_0+12\right) {r_0}^8\right.\vspace{1.5ex}\\
&\left.-6 \beta A_1^2 B_0^2 \left(6 A_1^4 B_0^4-24 A_1^3 B_0^3-57 A_1^2 B_0^2-42 A_1 B_0-8\right) {r_0}^4\right.\vspace{1.5ex}\\
&\left.+9 \beta ^2 A_1^4 B_0^4 \left(21 A_1^4 B_0^4+75 A_1^3 B_0^3+87 A_1^2 B_0^2+41 A_1 B_0+6\right)\right)\Biggr\}
\end{array}
\end{equation} 

\begin{align}
    A_4=& \frac{1}{2880 A_1^5 B_0^6 \left(\left(3 L^2 \beta -1\right) A_1 B_0-1\right)^8}\Biggl\{891813888 r_0^{18} \beta ^5 A_1^5 B_0^5 L^{19} \nonumber\\
   &-5971968 r_0^{12} \beta ^5 A_1^5 B_0^5 \left(r_0^4 (26 A_1 B_0-224)-3 \beta  A_1^2 B_0^2 (33 A_1 B_0+40)\right) L^{18}\nonumber\\
    &-41472 r_0^6 \beta ^4 A_1^4 B_0^4 \left(32 (1109 A_1 B_0+1108) r_0^{12}-24 \beta  A_1 B_0 \left(35 A_1^2 B_0^2-156 A_1 B_0+672\right) r_0^8\right.\nonumber\\
    &\left.-36 \beta ^2 A_1^3 B_0^3 \left(15 A_1^2 B_0^2+466 A_1 B_0+480\right) r_0^4\right.\nonumber\\
    &\left.-27 \beta ^3 A_1^5 B_0^5 \left(201 A_1^2 B_0^2+358 A_1 B_0+288\right)\right) L^{17}\nonumber\\
    &+5184 \beta ^4 A_1^4 B_0^4 \left(64 \left(749 A_1^2 B_0^2-5978 A_1 B_0-6648\right) r_0^{16}\right.\nonumber\\
    &\left.-24 \beta  A_1 B_0 \left(9767 A_1^3 B_0^3+20248 A_1^2 B_0^2+11272 A_1 B_0-896\right) r_0^{12}\right.\nonumber\\
    &\left.-18 \beta ^2 A_1^3 B_0^3 \left(181 A_1^3 B_0^3+232 A_1^2 B_0^2-2144 A_1 B_0-1920\right) r_0^8\right.\nonumber\\
    &\left.-54 \beta ^3 A_1^5 B_0^5 \left(75 A_1^3 B_0^3-164 A_1^2 B_0^2-462 A_1 B_0-576\right) r_0^4\right.\nonumber\\
    &\left.+81 \beta ^4 A_1^7 B_0^7 \left(61 A_1^3 B_0^3-84 A_1^2 B_0^2-260 A_1 B_0-168\right)\right) L^{16}\nonumber\\
    &+1728 r_0^6 \beta ^3 A_1^3 B_0^3 \left(32 \left(17575 A_1^2 B_0^2+34964 A_1 B_0+16864\right) r_0^{12}\right.\nonumber\\
    &\left.-24 \beta  A_1 B_0 \left(1309 A_1^3 B_0^3-5084 A_1^2 B_0^2+21208 A_1 B_0+26592\right) r_0^8\right.\nonumber\\
    &\left.-36 \beta ^2 A_1^3 B_0^3 \left(611 A_1^3 B_0^3+22896 A_1^2 B_0^2+42456 A_1 B_0+21920\right) r_0^4\right.\nonumber 
    \end{align}
 
\begin{align}
    &\left.-27 \beta ^3 A_1^5 B_0^5 \left(11647 A_1^3 B_0^3+29660 A_1^2 B_0^2+30012 A_1 B_0+10224\right)\right) L^{15}\nonumber\\
    &-216 \beta ^3 A_1^3 B_0^3 \left(128 \left(5740 A_1^3 B_0^3-42811 A_1^2 B_0^2-101468 A_1 B_0-50592\right) r_0^{16}\right.\nonumber\\
    &\left.-192 \beta  A_1 B_0 \left(25148 A_1^4 B_0^4+74151 A_1^3 B_0^3+76846 A_1^2 B_0^2+23300 A_1 B_0-4432\right) r_0^{12}\right.\nonumber\\
    &\left.-72 \beta ^2 A_1^3 B_0^3 \left(2263 A_1^4 B_0^4+4289 A_1^3 B_0^3-24448 A_1^2 B_0^2-44136 A_1 B_0-21920\right) r_0^8\right.\nonumber\\
    &\left.-324 \beta ^3 A_1^5 B_0^5 \left(757 A_1^4 B_0^4-1268 A_1^3 B_0^3-6766 A_1^2 B_0^2-9244 A_1 B_0-3408\right) r_0^4\right.\nonumber\\
    &\left.+81 \beta ^4 A_1^7 B_0^7 \left(4009 A_1^4 B_0^4-1277 A_1^3 B_0^3-18230 A_1^2 B_0^2-18420 A_1 B_0-4240\right)\right) L^{14}\nonumber\\
    &-144 r_0^6 \beta ^2 A_1^2 B_0^2 \left(128 \left(17445 A_1^3 B_0^3+51924 A_1^2 B_0^2+49948 A_1 B_0+15472\right) r_0^{12}\right.\nonumber\\
    &\left.-96 \beta  A_1 B_0 \left(2436 A_1^4 B_0^4-8220 A_1^3 B_0^3+33313 A_1^2 B_0^2+98044 A_1 B_0+50592\right) r_0^8\right.\nonumber\\
    &\left.-36 \beta ^2 A_1^3 B_0^3 \left(4955 A_1^4 B_0^4+234088 A_1^3 B_0^3+632164 A_1^2 B_0^2+610608 A_1 B_0+211072\right) r_0^4\right.\nonumber\\
    &\left.-27 \beta ^3 A_1^5 B_0^5 \left(144887 A_1^4 B_0^4+483300 A_1^3 B_0^3+634826 A_1^2 B_0^2+367064 A_1 B_0+72000\right)\right) L^{13}\nonumber\\
    &+18 \beta ^2 A_1^2 B_0^2 \left(512 \left(5484 A_1^4 B_0^4-38677 A_1^3 B_0^3-146716 A_1^2 B_0^2-149004 A_1 B_0-46416\right) r_0^{16}\right.\nonumber\\
    &\left.-192 \beta  A_1 B_0 \left(138699 A_1^5 B_0^5+532871 A_1^4 B_0^4\right.\right.\nonumber\\
    &\left.\left.+795648 A_1^3 B_0^3+505582 A_1^2 B_0^2+66424 A_1 B_0-33728\right) r_0^{12}\right.\nonumber\\
    &\left.-144 \beta ^2 A_1^3 B_0^3 \left(11770 A_1^5 B_0^5+29857 A_1^4 B_0^4\right.\right.\nonumber\\
    &\left.\left.-110436 A_1^3 B_0^3-329450 A_1^2 B_0^2-307400 A_1 B_0-105536\right) r_0^8\right.\nonumber\\
    &\left.-108 \beta ^3 A_1^5 B_0^5 \left(29377 A_1^5 B_0^5-33804 A_1^4 B_0^4-337692 A_1^3 B_0^3-577166 A_1^2 B_0^2-362040 A_1 B_0\right.\right.\nonumber\\
    &\left.\left.-72000\right) r_0^4+81 \beta ^4 A_1^7 B_0^7 \left(57749 A_1^5 B_0^5+43531 A_1^4 B_0^4-217888 A_1^3 B_0^3\right.\right.\nonumber\\
    &\left.\left.-360546 A_1^2 B_0^2-172264 A_1 B_0-27456\right)\right) L^{12}\nonumber\\
   &+24 r_0^6 \beta  A_1 B_0 \left(128 \left(17386 A_1^4 B_0^4+69063 A_1^3 B_0^3+100482 A_1^2 B_0^2+64648 A_1 B_0+16320\right) r_0^{12}\right.\nonumber\\
    &\left.-384 \beta  A_1 B_0 \left(1127 A_1^5 B_0^5-3318 A_1^4 B_0^4+12928 A_1^3 B_0^3+68448 A_1^2 B_0^2+74082 A_1 B_0+23208\right) r_0^8\right.\nonumber\\
    &\left.-144 \beta ^2 A_1^3 B_0^3 \left(2490 A_1^5 B_0^5+159306 A_1^4 B_0^4\right.\right.\nonumber\\
    &\left.\left.+570715 A_1^3 B_0^3+803940 A_1^2 B_0^2+524024 A_1 B_0+129504\right) r_0^4\right.\nonumber\\
    &\left.-27 \beta ^3 A_1^5 B_0^5 \left(502259 A_1^5 B_0^5+2084564 A_1^4 B_0^4+3461556 A_1^3 B_0^3\right.\right.\nonumber\\
    &\left.\left.+2775856 A_1^2 B_0^2+1022544 A_1 B_0+148544\right)\right) L^{11}\nonumber\\
      &-3 \beta  A_1 B_0 \left(512 \left(5224 A_1^5 B_0^5-35378 A_1^4 B_0^4-190741 A_1^3 B_0^3\right.\right.\nonumber\\
    &\left.\left.-297058 A_1^2 B_0^2-192792 A_1 B_0-48960\right) r_0^{16}\right.\nonumber\\
    &\left.-384 \beta  A_1 B_0 \left(108185 A_1^6 B_0^6+514663 A_1^5 B_0^5+1001953 A_1^4 B_0^4\right.\right.\nonumber\\
    &\left.\left.+950766 A_1^3 B_0^3+368616 A_1^2 B_0^2-21096 A_1 B_0-30944\right) r_0^{12}\right.\nonumber\\
     &\left.-288 \beta ^2 A_1^3 B_0^3 \left(16300 A_1^6 B_0^6+52208 A_1^5 B_0^5-122595 A_1^4 B_0^4\right.\right.\nonumber
    \end{align}
\begin{align}   
    &\left.\left.-583985 A_1^3 B_0^3-815324 A_1^2 B_0^2-526312 A_1 B_0-129504\right) r_0^8\right.\nonumber\\
    &\left.-108 \beta ^3 A_1^5 B_0^5 \left(105250 A_1^6 B_0^6-65163 A_1^5 B_0^5-1456552 A_1^4 B_0^4\right.\right.\nonumber\\
    &\left.\left.-3164156 A_1^3 B_0^3-2774160 A_1^2 B_0^2-1040688 A_1 B_0-148544\right) r_0^4\right.\nonumber\\
    &\left.+81 \beta ^4 A_1^7 B_0^7 \left(238435 A_1^6 B_0^6+440055 A_1^5 B_0^5-453246 A_1^4 B_0^4\right.\right.\nonumber\\
    &\left.\left.-1597446 A_1^3 B_0^3-1279464 A_1^2 B_0^2-411216 A_1 B_0-43456\right)\right) L^{10}\nonumber\\
    &-48 r_0^6 \left(64 \left(1162 A_1^5 B_0^5+5805 A_1^4 B_0^4+11508 A_1^3 B_0^3+11860 A_1^2 B_0^2+7080 A_1 B_0+2400\right) r_0^{12}\right.\nonumber\\
    &\left.-16 \beta  A_1 B_0 \left(2072 A_1^6 B_0^6-5416 A_1^5 B_0^5+19450 A_1^4 B_0^4+171097 A_1^3 B_0^3\right.\right.\nonumber\\
    &\left.\left.+291582 A_1^2 B_0^2+191640 A_1 B_0+48960\right) r_0^8\right.\nonumber\\
    &\left.-24 \beta ^2 A_1^3 B_0^3 \left(1225 A_1^6 B_0^6+121734 A_1^5 B_0^5+549834 A_1^4 B_0^4\right.\right.\nonumber\\
    &\left.\left.+1024407 A_1^3 B_0^3+973526 A_1^2 B_0^2+457784 A_1 B_0+77120\right) r_0^4\right.\nonumber\\
    &\left.-9 \beta ^3 A_1^5 B_0^5 \left(262288 A_1^6 B_0^6+1309168 A_1^5 B_0^5+2665786 A_1^4 B_0^4\right.\right.\nonumber\\
    &\left.\left.+2758629 A_1^3 B_0^3+1478542 A_1^2 B_0^2+389976 A_1 B_0+37696\right)\right) L^9\nonumber\\
    &+\left(512 \left(992 A_1^6 B_0^6-6574 A_1^5 B_0^5-47061 A_1^4 B_0^4-100384 A_1^3 B_0^3\right.\right.\nonumber\\
    &\left.\left.-103684 A_1^2 B_0^2-61560 A_1 B_0-21600\right) r_0^{16}\right.\nonumber\\
    &\left.-384 \beta  A_1 B_0 \left(45317 A_1^7 B_0^7+258249 A_1^6 B_0^6+621826 A_1^5 B_0^5\right.\right.\nonumber\\
    &\left.\left.+782965 A_1^4 B_0^4+491757 A_1^3 B_0^3+88426 A_1^2 B_0^2-39784 A_1 B_0-16320\right) r_0^{12}\right.\nonumber\\
    &\left.-288 \beta ^2 A_1^3 B_0^3 \left(12680 A_1^7 B_0^7+49342 A_1^6 B_0^6-65489 A_1^5 B_0^5\right.\right.\nonumber\\
    &\left.\left.-541303 A_1^4 B_0^4-1039649 A_1^3 B_0^3-986034 A_1^2 B_0^2-462744 A_1 B_0-77120\right) r_0^8\right.\nonumber\\
    &\left.-216 \beta ^3 A_1^5 B_0^5 \left(56410 A_1^7 B_0^7-4766 A_1^6 B_0^6-901913 A_1^5 B_0^5\right.\right.\nonumber\\
    &\left.\left.-2459528 A_1^4 B_0^4-2817077 A_1^3 B_0^3-1556554 A_1^2 B_0^2-406680 A_1 B_0-37696\right) r_0^4\right.\nonumber\\
    &\left.+81 \beta ^4 A_1^7 B_0^7 \left(308945 A_1^7 B_0^7+914075 A_1^6 B_0^6+342886 A_1^5 B_0^5\right.\right.\nonumber\\
    &\left.\left.-1512538 A_1^4 B_0^4-2124802 A_1^3 B_0^3-1129076 A_1^2 B_0^2-255472 A_1 B_0-19072\right)\right) L^8\nonumber\\
     &-24 r_0^6 \left(32 \left(126 A_1^7 B_0^7-304 A_1^6 B_0^6+900 A_1^5 B_0^5+13403 A_1^4 B_0^4\right.\right.\nonumber\\
    &\left.\left.+32076 A_1^3 B_0^3+33548 A_1^2 B_0^2+19800 A_1 B_0+7200\right) r_0^8\right.\nonumber\\
    &\left.+16 \beta  A_1^2 B_0^2 \left(220 A_1^7 B_0^7+49502 A_1^6 B_0^6+273371 A_1^5 B_0^5\right.\right.\nonumber\\
    &\left.\left.+639529 A_1^4 B_0^4+803617 A_1^3 B_0^3+555864 A_1^2 B_0^2+187736 A_1 B_0+23712\right) r_0^4\right.\nonumber\\
    &\left.+3 \beta ^2 A_1^4 B_0^4 \left(330418 A_1^7 B_0^7+1935576 A_1^6 B_0^6+4709353 A_1^5 B_0^5\right.\right.\nonumber\\
    &\left.\left.+6032726 A_1^4 B_0^4+4276472 A_1^3 B_0^3+1641704 A_1^2 B_0^2+305232 A_1 B_0+20928\right)\right) L^7\nonumber\\
      &+\left(128 \left(7971 A_1^8 B_0^8+53165 A_1^7 B_0^7+153225 A_1^6 B_0^6+240997 A_1^5 B_0^5\right.\right.\nonumber\\
    &\left.\left.+207437 A_1^4 B_0^4+76308 A_1^3 B_0^3-10988 A_1^2 B_0^2-19080 A_1 B_0-7200\right) r_0^{12}\right.\nonumber
    \end{align}
 
\begin{align}    
    &\left.+96 \beta  A_1^2 B_0^2 \left(5254 A_1^8 B_0^8+24166 A_1^7 B_0^7-11913 A_1^6 B_0^6\right.\right.\nonumber\\
    &\left.\left.-253171 A_1^5 B_0^5-643372 A_1^4 B_0^4-818867 A_1^3 B_0^3-569832 A_1^2 B_0^2-192040 A_1 B_0-23712\right) r_0^8\right.\nonumber\\
    &\left.+36 \beta ^2 A_1^4 B_0^4 \left(72380 A_1^8 B_0^8+32574 A_1^7 B_0^7-1300710 A_1^6 B_0^6\right.\right.\nonumber\\
    &\left.\left.-4376623 A_1^5 B_0^5-6291458 A_1^4 B_0^4-4661792 A_1^3 B_0^3-1800328 A_1^2 B_0^2-325744 A_1 B_0-20928\right) r_0^4\right.\nonumber\\
    &\left.-27 \beta ^3 A_1^6 B_0^6 \left(257494 A_1^8 B_0^8+1053494 A_1^7 B_0^7+1361749 A_1^6 B_0^6\right.\right.\nonumber\\
    &\left.\left.+68595 A_1^5 B_0^5-1349132 A_1^4 B_0^4-1271758 A_1^3 B_0^3-490352 A_1^2 B_0^2-77520 A_1 B_0-2880\right)\right) L^6\nonumber\\
    &+16 r_0^6 A_1^2 B_0^2 (A_1 B_0+1) \left(4 \left(-21 A_1^6 B_0^6+16753 A_1^5 B_0^5+94051 A_1^4 B_0^4\right.\right.\nonumber\\
    &\left.\left.+221145 A_1^3 B_0^3+276244 A_1^2 B_0^2+180688 A_1 B_0+46656\right) r_0^4\right.\nonumber\\
    &\left.+3 \beta  A_1^2 B_0^2 \left(58141 A_1^6 B_0^6+334207 A_1^5 B_0^5+782546 A_1^4 B_0^4\right.\right.\nonumber\\
    &\left.\left.+936061 A_1^3 B_0^3+590704 A_1^2 B_0^2+186008 A_1 B_0+22752\right)\right) L^5\nonumber\\
    &+A_1^2 B_0^2 (A_1 B_0+1) \left(-32 \left(906 A_1^7 B_0^7+3919 A_1^6 B_0^6-2822 A_1^5 B_0^5\right.\right.\nonumber\\
    &\left.\left.-44173 A_1^4 B_0^4-111314 A_1^3 B_0^3-142304 A_1^2 B_0^2-95260 A_1 B_0-25104\right) r_0^8\right.\nonumber\\
    &\left.-24 \beta  A_1^2 B_0^2 \left(12869 A_1^7 B_0^7-231 A_1^6 B_0^6-254792 A_1^5 B_0^5-786785 A_1^4 B_0^4\right.\right.\nonumber\\
    &\left.\left.-1037178 A_1^3 B_0^3-678848 A_1^2 B_0^2-214240 A_1 B_0-25344\right) r_0^4\right.\nonumber\\
    &\left.+9 \beta ^2 A_1^4 B_0^4 \left(134892 A_1^7 B_0^7+571752 A_1^6 B_0^6+863639 A_1^5 B_0^5\right.\right.\nonumber\\
    &\left.\left.+465790 A_1^4 B_0^4-102004 A_1^3 B_0^3-213928 A_1^2 B_0^2-78008 A_1 B_0-7840\right)\right) L^4\nonumber\\
    &-8 r_0^6 A_1^4 B_0^4 (A_1 B_0+1)^2 \left(17643 A_1^5 B_0^5+99838 A_1^4 B_0^4+224838 A_1^3 B_0^3  \right.\nonumber\\
    &\left.+248712 A_1^2 B_0^2+133336 A_1 B_0+27744\right) L^3\nonumber\\
    &+15 A_1^6 B_0^6 (A_1 B_0+1)^3 \left(359 A_1^5 B_0^5+1632 A_1^4 B_0^4+2908 A_1^3 B_0^3+2544 A_1^2 B_0^2+1104 A_1 B_0+192\right)\nonumber\\
    &-A_1^4 B_0^4 (A_1 B_0 L+L)^2 \left(4 \left(-3914 A_1^6 B_0^6+1919 A_1^5 B_0^5+84470 A_1^4 B_0^4\right.\right.\nonumber\\
    &\left.\left.+241480 A_1^3 B_0^3+291320 A_1^2 B_0^2+162560 A_1 B_0+34560\right) r_0^4\right.\nonumber\\
    &\left.+3 \beta  A_1^2 B_0^2 \left(40623 A_1^6 B_0^6+178461 A_1^5 B_0^5+298868 A_1^4 B_0^4  \right.\right.\nonumber\\
    &\left.\left.+232906 A_1^3 B_0^3+78672 A_1^2 B_0^2+3464 A_1 B_0-2720\right)\right) \Biggr\}
    \end{align}
 
\begin{align}
  B_3&= -\frac{L^2 }{180 A_1^6 B_0^5 \left(\left(3 L^2 \beta -1\right) A_1 B_0-1\right)^7}\Biggl\{9289728 L^{15} \beta ^4 A_1^4 B_0^4 r_0^{18}\nonumber\\
    &-497664 L^{14} \beta ^4 A_1^4 B_0^4 \left(4 (2 A_1 B_0-7) r_0^4+\beta  A_1^2 B_0^2 (19-6 A_1 B_0)\right) r_0^{12}\nonumber\\
    &-8 L A_1^4 B_0^4 (A_1 B_0+1)^2 \left(63 A_1^4 B_0^4+378 A_1^3 B_0^3+602 A_1^2 B_0^2+384 A_1 B_0+88\right) r_0^6\nonumber\\
    &-3456 L^{13} \beta ^3 A_1^3 B_0^3 \left(16 (237 A_1 B_0+296) r_0^{12}-36 \beta  A_1 B_0 \left(3 A_1^2 B_0^2-32 A_1 B_0+56\right) r_0^8\right.\nonumber
    \end{align}
 
\begin{align}
    &\left.+9 \beta ^2 A_1^3 B_0^3 \left(45 A_1^2 B_0^2-136 A_1 B_0+304\right) r_0^4-27 \beta ^3 A_1^5 B_0^5 \left(3 A_1^2 B_0^2+16 A_1 B_0+30\right)\right) r_0^6\nonumber\\
    &+144 L^{11} \beta ^2 A_1^2 B_0^2 \left(512 \left(93 A_1^2 B_0^2+224 A_1 B_0+129\right) r_0^{12}\right.\nonumber\\
    &\left.-288 \beta  A_1 B_0 \left(12 A_1^3 B_0^3-121 A_1^2 B_0^2+117 A_1 B_0+296\right) r_0^8\right.\nonumber\\
    &\left.+72 \beta ^2 A_1^3 B_0^3 \left(225 A_1^3 B_0^3-405 A_1^2 B_0^2+758 A_1 B_0+1072\right) r_0^4\right.\nonumber\\
    &\left.-27 \beta ^3 A_1^5 B_0^5 \left(69 A_1^3 B_0^3+572 A_1^2 B_0^2+1998 A_1 B_0+624\right)\right) r_0^6\nonumber\\
    &-24 L^9 \beta  A_1 B_0 \left(128 \left(514 A_1^3 B_0^3+1797 A_1^2 B_0^2+1986 A_1 B_0+592\right) r_0^{12}\right.\nonumber\\
    &\left.-192 \beta  A_1 B_0 \left(54 A_1^4 B_0^4-513 A_1^3 B_0^3+70 A_1^2 B_0^2+2324 A_1 B_0+1548\right) r_0^8\right.\nonumber\\
    &\left.+144 \beta ^2 A_1^3 B_0^3 \left(450 A_1^4 B_0^4-255 A_1^3 B_0^3+727 A_1^2 B_0^2+2868 A_1 B_0+1032\right) r_0^4\right.\nonumber\\
    &\left.-27 \beta ^3 A_1^5 B_0^5 \left(9 A_1^4 B_0^4+1362 A_1^3 B_0^3+9700 A_1^2 B_0^2+7744 A_1 B_0+2096\right)\right) r_0^6\nonumber\\
    &+48 L^3 A_1^2 B_0^2 (A_1 B_0+1) \left(4 \left(30 A_1^5 B_0^5+66 A_1^4 B_0^4+118 A_1^3 B_0^3+297 A_1^2 B_0^2+276 A_1 B_0\right.\right.\nonumber\\
    &\left.\left.+80\right) r_0^4+\beta  A_1^2 B_0^2 \left(153 A_1^5 B_0^5+870 A_1^4 B_0^4+1036 A_1^3 B_0^3+309 A_1^2 B_0^2-132 A_1 B_0-56\right)\right) r_0^6\nonumber\\
    &+48 L^7 \left(64 \left(44 A_1^4 B_0^4+199 A_1^3 B_0^3+318 A_1^2 B_0^2+188 A_1 B_0+40\right) r_0^{12}\right.\nonumber\\
    &\left.-16 \beta  A_1 B_0 \left(72 A_1^5 B_0^5-642 A_1^4 B_0^4-482 A_1^3 B_0^3+3979 A_1^2 B_0^2+5742 A_1 B_0+1776\right) r_0^8\right.\nonumber\\
    &\left.+24 \beta ^2 A_1^3 B_0^3 \left(450 A_1^5 B_0^5+305 A_1^4 B_0^4+672 A_1^3 B_0^3+3174 A_1^2 B_0^2+2478 A_1 B_0+624\right) r_0^4\right.\nonumber\\
    &\left.+9 \beta ^3 A_1^5 B_0^5 \left(192 A_1^5 B_0^5+361 A_1^4 B_0^4-4181 A_1^3 B_0^3-6493 A_1^2 B_0^2-3478 A_1 B_0-496\right)\right) r_0^6\nonumber\\
    &+8 L^5 \left(32 \left(18 A_1^6 B_0^6-150 A_1^5 B_0^5-256 A_1^4 B_0^4\right.\right.\nonumber\\
    &\left.+1107 A_1^3 B_0^3+2646 A_1^2 B_0^2+1676 A_1 B_0+360\right) r_0^8\nonumber\\
    &-24 \beta  A_1^2 B_0^2 \left(450 A_1^6 B_0^6+870 A_1^5 B_0^5+1350 A_1^4 B_0^4\right.\nonumber\\
    &\left.+4001 A_1^3 B_0^3+4632 A_1^2 B_0^2+2128 A_1 B_0+240\right) r_0^4\nonumber\\
    &-9 \beta ^2 A_1^4 B_0^4 \left(546 A_1^6 B_0^6+2738 A_1^5 B_0^5+213 A_1^4 B_0^4\right.\nonumber\\
    &\left.\left.-5766 A_1^3 B_0^3-5964 A_1^2 B_0^2-1928 A_1 B_0-144\right) \right)r_0^6\nonumber\\
    &+1296 L^{12} \beta ^3 A_1^3 B_0^3 \left(32 \left(127 A_1^2 B_0^2-354 A_1 B_0-592\right) r_0^{16}\right.\nonumber\\
    &\left.+16 \beta  A_1 B_0 \left(-243 A_1^3 B_0^3+386 A_1^2 B_0^2+488 A_1 B_0+56\right) r_0^{12}\right.\nonumber\\
    &\left.+12 \beta ^2 A_1^3 B_0^3 \left(9 A_1^3 B_0^3-40 A_1^2 B_0^2+88 A_1 B_0-152\right) r_0^8\right.\nonumber\\
    &\left.+9 \beta ^3 A_1^5 B_0^5 \left(3 A_1^3 B_0^3-A_1^2 B_0^2+48 A_1 B_0+120\right) r_0^4\right.\nonumber\\
    &\left.+27 \beta ^4 A_1^7 B_0^7 \left(5 A_1^3 B_0^3+19 A_1^2 B_0^2-4 A_1 B_0-14\right)\right)\nonumber\\
    &-18 L^{10} \beta ^2 A_1^2 B_0^2 \left(256 \left(567 A_1^3 B_0^3-1177 A_1^2 B_0^2-5012 A_1 B_0-3096\right) r_0^{16}\right.\nonumber\\
     &\left.-96 \beta  A_1 B_0 \left(1965 A_1^4 B_0^4-328 A_1^3 B_0^3-4686 A_1^2 B_0^2-2292 A_1 B_0-1184\right) r_0^{12}\right.\nonumber\\
    &\left.+288 \beta ^2 A_1^3 B_0^3 \left(45 A_1^4 B_0^4-146 A_1^3 B_0^3+190 A_1^2 B_0^2-381 A_1 B_0-536\right) r_0^8\right.\nonumber\\
        &\left.+108 \beta ^3 A_1^5 B_0^5 \left(33 A_1^4 B_0^4+11 A_1^3 B_0^3+455 A_1^2 B_0^2+1886 A_1 B_0+624\right) r_0^4\right.\nonumber
    \end{align}
 
\begin{align}
    &\left.+81 \beta ^4 A_1^7 B_0^7 \left(225 A_1^4 B_0^4+969 A_1^3 B_0^3+446 A_1^2 B_0^2-230 A_1 B_0-272\right)\right)\nonumber\\
     &+A_1^5 B_0^5 (A_1 B_0+1)^2 \left(4 \left(6 A_1^4 B_0^4+23 A_1^3 B_0^3-56 A_1^2 B_0^2-104 A_1 B_0-40\right) r_0^4\right.\nonumber\\
    &\left.+3 \beta  A_1 B_0 \left(50 A_1^5 B_0^5+223 A_1^4 B_0^4+206 A_1^3 B_0^3-94 A_1^2 B_0^2-208 A_1 B_0-72\right)\right)\nonumber\\
    &+3 L^8 \beta  A_1 B_0 \left(512 \left(375 A_1^4 B_0^4-512 A_1^3 B_0^3-4685 A_1^2 B_0^2-5850 A_1 B_0-1776\right) r_0^{16}\right.\nonumber\\
    &\left.-192 \beta  A_1 B_0 \left(1983 A_1^5 B_0^5+2233 A_1^4 B_0^4-2814 A_1^3 B_0^3-3722 A_1^2 B_0^2-4104 A_1 B_0-2064\right) r_0^{12}\right.\nonumber\\
    &\left.+144 \beta ^2 A_1^3 B_0^3 \left(360 A_1^5 B_0^5-736 A_1^4 B_0^4+7 A_1^3 B_0^3-2440 A_1^2 B_0^2-6096 A_1 B_0-2064\right) r_0^8\right.\nonumber\\
    &\left.+108 \beta ^3 A_1^5 B_0^5 \left(150 A_1^5 B_0^5+163 A_1^4 B_0^4+1618 A_1^3 B_0^3+9462 A_1^2 B_0^2+7656 A_1 B_0+2096\right) r_0^4\right.\nonumber\\
    &\left.+81 \beta ^4 A_1^7 B_0^7 \left(1050 A_1^5 B_0^5+5035 A_1^4 B_0^4+4985 A_1^3 B_0^3+1740 A_1^2 B_0^2-416 A_1 B_0+80\right)\right)\nonumber\\
    &-L^2 A_1^2 B_0^2 (A_1 B_0+1) \left(16 \left(36 A_1^6 B_0^6+20 A_1^5 B_0^5-181 A_1^4 B_0^4\right.\right.\nonumber\\
    &\left.\left.-824 A_1^3 B_0^3-1724 A_1^2 B_0^2-1368 A_1 B_0-360\right) r_0^8\right.\nonumber\\
    &\left.+24 \beta  A_1^2 B_0^2 \left(21 A_1^6 B_0^6+68 A_1^5 B_0^5-54 A_1^4 B_0^4+183 A_1^3 B_0^3+454 A_1^2 B_0^2+268 A_1 B_0+32\right) r_0^4\right.\nonumber\\
    &\left.+9 \beta ^2 A_1^4 B_0^4 \left(330 A_1^6 B_0^6+1684 A_1^5 B_0^5+2347 A_1^4 B_0^4+1052 A_1^3 B_0^3-176 A_1^2 B_0^2-152 A_1 B_0+8\right)\right)\nonumber\\
    &-L^6 \left(512 \left(93 A_1^5 B_0^5-61 A_1^4 B_0^4-1449 A_1^3 B_0^3-2754 A_1^2 B_0^2-1684 A_1 B_0-360\right) r_0^{16}\right.\nonumber\\
    &\left.-192 \beta  A_1 B_0 \left(999 A_1^6 B_0^6+2314 A_1^5 B_0^5+688 A_1^4 B_0^4\right.\right.\nonumber\\
    &\left.\left.-1410 A_1^3 B_0^3-3814 A_1^2 B_0^2-3996 A_1 B_0-1184\right) r_0^{12}\right.\nonumber\\
    &\left.+144 \beta ^2 A_1^3 B_0^3 \left(360 A_1^6 B_0^6-304 A_1^5 B_0^5-1019 A_1^4 B_0^4\right.\right.\nonumber\\
    &\left.\left.-3745 A_1^3 B_0^3-8620 A_1^2 B_0^2-5604 A_1 B_0-1248\right) r_0^8\right.\nonumber\\
    &\left.+216 \beta ^3 A_1^5 B_0^5 \left(90 A_1^6 B_0^6+176 A_1^5 B_0^5+655 A_1^4 B_0^4\right.\right.\nonumber\\
    &\left.\left.+5086 A_1^3 B_0^3+6755 A_1^2 B_0^2+3498 A_1 B_0+496\right) r_0^4\right.\nonumber\\
    &\left.+81 \beta ^4 A_1^7 B_0^7 \left(1300 A_1^6 B_0^6+6840 A_1^5 B_0^5+9665 A_1^4 B_0^4\right.\right.\nonumber\\
    &\left.\left.+6281 A_1^3 B_0^3+2056 A_1^2 B_0^2+996 A_1 B_0+352\right)\right)\nonumber\\
    &+L^4 \left(-64 \left(201 A_1^7 B_0^7+689 A_1^6 B_0^6+768 A_1^5 B_0^5+244 A_1^4 B_0^4\right.\right.\nonumber\\
    &\left.\left.-886 A_1^3 B_0^3-1812 A_1^2 B_0^2-1112 A_1 B_0-240\right) r_0^{12}\right.\nonumber\\
    &\left.+48 \beta  A_1^2 B_0^2 \left(180 A_1^7 B_0^7+64 A_1^6 B_0^6-779 A_1^5 B_0^5-3226 A_1^4 B_0^4\right.\right.\nonumber\\
    &\left.\left.-7216 A_1^3 B_0^3-6710 A_1^2 B_0^2-2624 A_1 B_0-240\right) r_0^8\right.\nonumber\\
    &\left.+36 \beta ^2 A_1^4 B_0^4 \left(120 A_1^7 B_0^7+358 A_1^6 B_0^6+442 A_1^5 B_0^5+4453 A_1^4 B_0^4\right.\right.\nonumber\\
    &\left.\left.+8680 A_1^3 B_0^3+6528 A_1^2 B_0^2+1848 A_1 B_0+144\right) r_0^4\right.\nonumber\\
    &\left.+27 \beta ^3 A_1^6 B_0^6 \left(900 A_1^7 B_0^7+5130 A_1^6 B_0^6+8965 A_1^5 B_0^5\right.\right.\nonumber\\
    &\left.\left.+7163 A_1^4 B_0^4+2994 A_1^3 B_0^3+1170 A_1^2 B_0^2+560 A_1 B_0+112\right)\right)\Biggr\}
    \end{align}

The coefficients $B_4$ and $A_5$ require about  5 and 10 pages, respectively, to present, so we omit them here.

 



\begin{thebibliography}{0}%
\makeatletter
\providecommand \@ifxundefined [1]{%
 \@ifx{#1\undefined}
}%
\providecommand \@ifnum [1]{%
 \ifnum #1\expandafter \@firstoftwo
 \else \expandafter \@secondoftwo
 \fi
}%
\providecommand \@ifx [1]{%
 \ifx #1\expandafter \@firstoftwo
 \else \expandafter \@secondoftwo
 \fi
}%
\providecommand \natexlab [1]{#1}%
\providecommand \enquote  [1]{``#1''}%
\providecommand \bibnamefont  [1]{#1}%
\providecommand \bibfnamefont [1]{#1}%
\providecommand \citenamefont [1]{#1}%
\providecommand \href@noop [0]{\@secondoftwo}%
\providecommand \href [0]{\begingroup \@sanitize@url \@href}%
\providecommand \@href[1]{\@@startlink{#1}\@@href}%
\providecommand \@@href[1]{\endgroup#1\@@endlink}%
\providecommand \@sanitize@url [0]{\catcode `\\12\catcode `\$12\catcode `\&12\catcode `\#12\catcode `\^12\catcode `\_12\catcode `\%12\relax}%
\providecommand \@@startlink[1]{}%
\providecommand \@@endlink[0]{}%
\providecommand \url  [0]{\begingroup\@sanitize@url \@url }%
\providecommand \@url [1]{\endgroup\@href {#1}{\urlprefix }}%
\providecommand \urlprefix  [0]{URL }%
\providecommand \Eprint [0]{\href }%
\providecommand \doibase [0]{http://dx.doi.org/}%
\providecommand \selectlanguage [0]{\@gobble}%
\providecommand \bibinfo  [0]{\@secondoftwo}%
\providecommand \bibfield  [0]{\@secondoftwo}%
\providecommand \translation [1]{[#1]}%
\providecommand \BibitemOpen [0]{}%
\providecommand \bibitemStop [0]{}%
\providecommand \bibitemNoStop [0]{.\EOS\space}%
\providecommand \EOS [0]{\spacefactor3000\relax}%
\providecommand \BibitemShut  [1]{\csname bibitem#1\endcsname}%
\let\auto@bib@innerbib\@empty
\end{thebibliography}%


\begin{thebibliography}{999}

\bibitem[Morris et~al.(1988)Morris, Thorne, and Yurtsever]{Morris:1988tu}
Morris, M.S.; Thorne, K.S.; Yurtsever, U.
\newblock {Wormholes, Time Machines, and the Weak Energy Condition}.
\newblock {\em Phys. Rev. Lett.} {\bf 1988}, {\em 61},~1446--1449.
\newblock {\url{https://doi.org/10.1103/PhysRevLett.61.1446}}.

\bibitem[Morris and Thorne(1988)]{Morris:1988cz}
Morris, M.S.; Thorne, K.S.
\newblock {Wormholes in space-time and their use for interstellar travel: A
  tool for teaching general relativity}.
\newblock {\em Am. J. Phys.} {\bf 1988}, {\em 56},~395--412.
\newblock {\url{https://doi.org/10.1119/1.15620}}.

\bibitem[Visser(1995)]{Visser:1995cc}
Visser, M.
\newblock {\em {Lorentzian Wormholes: From Einstein to Hawking}}; {American Institute of Physics}
: Woodbury, NY, USA, 1995.

\bibitem[Visser(1993)]{Visser:1992tx}
Visser, M.
\newblock {From wormhole to time machine: Comments on Hawking's chronology
  protection conjecture}.
\newblock {\em Phys. Rev. D} {\bf 1993}, {\em 47},~554--{565.} 
\newblock {\url{https://doi.org/10.1103/PhysRevD.47.554}}.

\bibitem[Ford and Roman(1996)]{Ford:1995wg}
Ford, L.H.; Roman, T.A.
\newblock {Quantum field theory constrains traversable wormhole geometries}.
\newblock {\em Phys. Rev. D} {\bf 1996}, {\em 53},~5496--5507.
\newblock {\url{https://doi.org/10.1103/PhysRevD.53.5496}}.

\bibitem[Harko et~al.(2013)Harko, Lobo, Mak, and Sushkov]{Harko:2013yb}
Harko, T.; Lobo, F.S.N.; Mak, M.K.; Sushkov, S.V.
\newblock {Modified-gravity wormholes without exotic matter}.
\newblock {\em Phys. Rev. D} {\bf 2013}, {\em 87},~067504.
\newblock {\url{https://doi.org/10.1103/PhysRevD.87.067504}}.

\bibitem[Nandi et~al.(1997)Nandi, Islam, and Evans]{Nandi:1997mx}
Nandi, K.K.; Islam, A.; Evans, J.
\newblock {Brans wormholes}.
\newblock {\em Phys. Rev. D} {\bf 1997}, {\em 55},~2497--2500.
\newblock {\url{https://doi.org/10.1103/PhysRevD.55.2497}}.

\bibitem[Yue and Gao(2011)]{Yue:2011cq}
Yue, X.; Gao, S.
\newblock {Stability of Brans-Dicke thin shell wormholes}.
\newblock {\em Phys. Lett. A} {\bf 2011}, {\em 375},~2193--2200.
\newblock {\url{https://doi.org/10.1016/j.physleta.2011.04.055}}.

\bibitem[Lobo and Oliveira(2010)]{Lobo:2010sb}
Lobo, F.S.N.; Oliveira, M.A.
\newblock {General class of vacuum Brans-Dicke wormholes}.
\newblock {\em Phys. Rev. D} {\bf 2010}, {\em 81},~067501.
\newblock {\url{https://doi.org/10.1103/PhysRevD.81.067501}}.

\bibitem[Sushkov and Kozyrev(2011)]{Sushkov:2011zh}
Sushkov, S.V.; Kozyrev, S.M.
\newblock {Composite vacuum Brans-Dicke wormholes}.
\newblock {\em Phys. Rev. D} {\bf 2011}, {\em 84},~124026.
\newblock {\url{https://doi.org/10.1103/PhysRevD.84.124026}}.

\bibitem[Lobo and Oliveira(2009)]{Lobo:2009ip}
Lobo, F.S.N.; Oliveira, M.A.
\newblock {Wormhole geometries in f(R) modified theories of gravity}.
\newblock {\em Phys. Rev. D} {\bf 2009}, {\em 80},~104012.
\newblock {\url{https://doi.org/10.1103/PhysRevD.80.104012}}.

\bibitem[Garcia and Lobo(2010)]{Garcia:2010xb}
Garcia, N.M.; Lobo, F.S.N.
\newblock {Wormhole geometries supported by a nonminimal curvature-matter
  coupling}.
\newblock {\em Phys. Rev. D} {\bf 2010}, {\em 82},~104018.
\newblock {\url{https://doi.org/10.1103/PhysRevD.82.104018}}.
 
\bibitem[Montelongo~Garcia and Lobo(2011)]{MontelongoGarcia:2010xd}
Montelongo~Garcia, N.; Lobo, F.S.N.
\newblock {Nonminimal curvature-matter coupled wormholes with matter satisfying
  the null energy condition}.
\newblock {\em Class. Quant. Grav.} {\bf 2011}, {\em 28},~085018.
\newblock {\url{https://doi.org/10.1088/0264-9381/28/8/085018}}.

\bibitem[Anabalon et~al.(2019)Anabalon, Oliva, and Quijada]{Anabalon:2019lzc}
Anabalon, A.; Oliva, J.; Quijada, C.
\newblock {Fully resonant scalars on asymptotically AdS wormholes}.
\newblock {\em Phys. Rev. D} {\bf 2019}, {\em 99},~104022.
\newblock {\url{https://doi.org/10.1103/PhysRevD.99.104022}}.

\bibitem[Lovelock(1970)]{Lovelock1}
Lovelock, D.
\newblock Divergence-free tensorial concomitants.
\newblock {\em Aequationes Math.} {\bf 1970}, {\em 4},~127--138.
\newblock {\url{https://doi.org/10.1007/BF01817753}}.

\bibitem[Lovelock(1971)]{Lovelock2}
Lovelock, D.
\newblock {The Einstein tensor and its generalizations}.
\newblock {\em J. Math. Phys.} {\bf 1971}, {\em 12},~498--501.
\newblock {\url{https://doi.org/10.1063/1.1665613}}.

\bibitem[Bhawal and Kar(1992)]{Bhawal:1992sz}
Bhawal, B.; Kar, S.~{Lorentzian wormholes in Einstein-Gauss-Bonnet theory}.~{\em Phys. Rev. D} {\bf 1992}, {\em 46},~2464--2468.
\newblock {\url{https://doi.org/10.1103/PhysRevD.46.2464}}.

\bibitem[Wang and Li(1996)]{Wang:1996xn}
Wang, W.; Li, X.Y.
\newblock {Wormhole solution in D-dimensional Einstein and Lovelock theories}.
\newblock {\em  Il Nuovo Cimento B} {\bf 1996}, {\em 111},~1101--1110.
\newblock {\url{https://doi.org/10.1007/BF02743221}}.

\bibitem[Shang and Xu(1999)]{Shang:1999xs}
Shang, Y.W.; Xu, J.J.
\newblock {Wormhole solution in Lovelock gravity theory}.
\newblock {\em Chin. Phys. Lett.} {\bf 1999}, {\em 16},~85--87.
\newblock {\url{https://doi.org/10.1088/0256-307X/16/2/003}}.

\bibitem[Maeda and Nozawa(2008)]{Maeda:2008nz}
Maeda, H.; Nozawa, M.
\newblock {Static and symmetric wormholes respecting energy conditions in
  Einstein-Gauss-Bonnet gravity}.
\newblock {\em Phys.~Rev. D} {\bf 2008}, {\em 78},~024005.
\newblock {\url{https://doi.org/10.1103/PhysRevD.78.024005}}.

\bibitem[Dehghani and Dayyani(2009)]{Dehghani:2009zza}
Dehghani, M.H.; Dayyani, Z.
\newblock {Lorentzian wormholes in Lovelock gravity}.
\newblock {\em Phys. Rev. D} {\bf 2009}, {\em 79},~064010.
\newblock {\url{https://doi.org/10.1103/PhysRevD.79.064010}}.

\bibitem[Mehdizadeh and Lobo(2016)]{Mehdizadeh:2016nna}
Mehdizadeh, M.R.; Lobo, F.S.N.
\newblock {Novel third-order Lovelock wormhole solutions}.
\newblock {\em Phys. Rev. D} {\bf 2016}, {\em 93},~124014.
\newblock {\url{https://doi.org/10.1103/PhysRevD.93.124014}}.

\bibitem[Mehdizadeh and Riazi(2012)]{Mehdizadeh:2012zz}
Mehdizadeh, M.R.; Riazi, N.
\newblock {Cosmological wormholes in Lovelock gravity}.
\newblock {\em Phys. Rev. D} {\bf 2012}, {\em 85},~124022.
\newblock {\url{https://doi.org/10.1103/PhysRevD.85.124022}}.

\bibitem[Mehdizadeh et~al.(2015)Mehdizadeh, Kord~Zangeneh, and
  Lobo]{Mehdizadeh:2015jra}
Mehdizadeh, M.R.; Kord~Zangeneh, M.; Lobo, F.S.N.
\newblock {Einstein-Gauss-Bonnet traversable wormholes satisfying the weak
  energy condition}.
\newblock {\em Phys. Rev. D} {\bf 2015}, {\em 91},~084004,
\newblock {\url{https://doi.org/10.1103/PhysRevD.91.084004}}.

\bibitem[Oliva and Ray(2010)]{Quasi2}
Oliva, J.; Ray, S.
\newblock {A new cubic theory of gravity in five dimensions: Black hole,
  Birkhoff's theorem and C-function}.
\newblock {\em Class. Quant. Grav.} {\bf 2010}, {\em 27},~225002.
\newblock {\url{https://doi.org/10.1088/0264-9381/27/22/225002}}.

\bibitem[Myers and Robinson(2010)]{Quasi}
{Myers, R.C.; Robinson, B.
\newblock {Black Holes in Quasi-topological Gravity}.
\newblock {\em JHEP} {\bf 2010}, {\em 08},~067.
\newblock {\url{https://doi.org/10.1007/JHEP08(2010)067}}.}

\bibitem[Oliva and Ray(2011)]{Oliva:2011xu}
Oliva, J.; Ray, S.
\newblock {Birkhoff's Theorem in Higher Derivative Theories of Gravity}.
\newblock {\em Class. Quant. Grav.} {\bf 2011}, {\em 28},~175007.
\newblock {\url{https://doi.org/10.1088/0264-9381/28/17/175007}}.

\bibitem[Oliva and Ray(2012)]{Oliva:2012zs}
Oliva, J.; Ray, S.
\newblock {Birkhoff's Theorem in Higher Derivative Theories of Gravity II}.
\newblock {\em Phys. Rev.} {\bf 2012}, {\em D86},~084014.
\newblock {\url{https://doi.org/10.1103/PhysRevD.86.084014}}.

\bibitem[Dehghani et~al.(2012)Dehghani, Bazrafshan, Mann, Mehdizadeh,
  Ghanaatian, and Vahidinia]{Dehghani:2011vu}
Dehghani, M.H.; Bazrafshan, A.; Mann, R.B.; Mehdizadeh, M.R.; Ghanaatian, M.;
  Vahidinia, M.H.
\newblock {Black Holes in Quartic Quasitopological Gravity}.
\newblock {\em Phys. Rev.} {\bf 2012}, {\em D85},~104009.
\newblock {\url{https://doi.org/10.1103/PhysRevD.85.104009}}.

\bibitem[Cisterna et~al.(2017)Cisterna, Guajardo, Hassaine, and
  Oliva]{Cisterna:2017umf}
Cisterna, A.; Guajardo, L.; Hassaine, M.; Oliva, J.
\newblock {Quintic quasi-topological gravity}.
\newblock {\em JHEP} {\bf 2017}, {\em 4},~66. 
\newblock {\url{https://doi.org/10.1007/JHEP04(2017)066}}.

\bibitem[Hennigar et~al.(2017)Hennigar, Kubiz\v{n}\'ak, and
  Mann]{Hennigar:2017ego}
Hennigar, R.A.; Kubiz\v{n}\'ak, D.; Mann, R.B.
\newblock {Generalized quasitopological gravity}.
\newblock {\em Phys. Rev. D} {\bf 2017}, {\em 95},~104042.
\newblock {\url{https://doi.org/10.1103/PhysRevD.95.104042}}.

\bibitem[Bueno and Cano(2017)]{PabloPablo3}
Bueno, P.; Cano, P.A.
\newblock {On black holes in higher-derivative gravities}.
\newblock {\em Class. Quant. Grav.} {\bf 2017}, {\em 34},~175008.
\newblock {\url{https://doi.org/10.1088/1361-6382/aa8056}}.

\bibitem[Ahmed et~al.(2017)Ahmed, Hennigar, Mann, and Mir]{Ahmed:2017jod}
Ahmed, J.; Hennigar, R.A.; Mann, R.B.; Mir, M.
\newblock {Quintessential Quartic Quasi-topological Quartet}.
\newblock {\em JHEP} {\bf 2017}, {\em 5},~134. 
\newblock {\url{https://doi.org/10.1007/JHEP05(2017)134}}.

\bibitem[Bueno and Cano(2017)]{PabloPablo4}
Bueno, P.; Cano, P.A.
\newblock {Universal black hole stability in four dimensions}.
\newblock {\em Phys. Rev.} {\bf 2017}, {\em D96},~024034.
\newblock {\url{https://doi.org/10.1103/PhysRevD.96.024034}}.

\bibitem[Bueno et~al.(2020)Bueno, Cano, and Hennigar]{Bueno:2019ycr}
Bueno, P.; Cano, P.A.; Hennigar, R.A.
\newblock {(Generalized) quasi-topological gravities at all orders}.
\newblock {\em Class. Quant. Grav.} {\bf 2020}, {\em 37},~015002.
\newblock {\url{https://doi.org/10.1088/1361-6382/ab5410}}.

\bibitem[Hennigar(2017)]{Hennigar:2017umz}
Hennigar, R.A.
\newblock {Criticality for charged black branes}.
\newblock {\em JHEP} {\bf 2017}, {\em 9},~82.
\newblock {\url{https://doi.org/10.1007/JHEP09(2017)082}}.

\bibitem[Mir and Mann(2019)]{Mir:2019rik}
Mir, M.; Mann, R.B.
\newblock {On generalized quasi-topological cubic-quartic gravity:
  Thermodynamics and holography}.
\newblock {\em JHEP} {\bf 2019}, {\em 7},~12. 
\newblock {\url{https://doi.org/10.1007/JHEP07(2019)012}}.

\bibitem[Mir et~al.(2019)Mir, Hennigar, Ahmed, and Mann]{Mir:2019ecg}
Mir, M.; Hennigar, R.A.; Ahmed, J.; Mann, R.B.
\newblock {Black hole chemistry and holography in generalized quasi-topological
  gravity}.
\newblock {\em JHEP} {\bf 2019}, {\em 08},~068.
\newblock {\url{https://doi.org/10.1007/JHEP08(2019)068}}.

\bibitem[Li et~al.(2018{\natexlab{a}})Li, Liu, and Lu]{Li:2017ncu}
{Li, Y.Z.; Liu, H.S.; Lu, H.
\newblock {Quasi-Topological Ricci Polynomial Gravities}.
\newblock {\em JHEP} {\bf 2018}, {\em 2},~166.}
\newblock {\url{https://doi.org/10.1007/JHEP02(2018)166}}.

\bibitem[Li et~al.(2018{\natexlab{b}})Li, Lu, and Wu]{Li:2017txk}
Li, Y.Z.; Lu, H.; Wu, J.B.
\newblock {Causality and a-theorem Constraints on Ricci Polynomial and Riemann
  Cubic Gravities}.
\newblock {\em Phys. Rev.} {\bf 2018}, {\em D97},~024023.
\newblock {\url{https://doi.org/10.1103/PhysRevD.97.024023}}.

\bibitem[Li et~al.(2018{\natexlab{c}})Li, Lu, and Mai]{Li:2018drw}
Li, Y.Z.; Lu, H.; Mai, Z.F.
\newblock {Universal Structure of Covariant Holographic Two-Point Functions In
  Massless Higher-Order Gravities}.
\newblock {\em JHEP} {\bf 2018}, {\em 10},~063.
\newblock {\url{https://doi.org/10.1007/JHEP10(2018)063}}.

\bibitem[Bueno et~al.(2019)Bueno, Cano, Moreno, and Murcia]{Bueno:2019ltp}
Bueno, P.; Cano, P.A.; Moreno, J.; Murcia, A.
\newblock {All higher-curvature gravities as Generalized quasi-topological
  gravities}.
\newblock {\em JHEP} {\bf 2019}, {\em 11},~062.
\newblock {\url{https://doi.org/10.1007/JHEP11(2019)062}}.

\bibitem[Bueno and Cano(2016)]{Bueno:2016xff}
Bueno, P.; Cano, P.A.
\newblock {Einsteinian cubic gravity}.
\newblock {\em Phys. Rev. D} {\bf 2016}, {\em 94},~104005.
\newblock {\url{https://doi.org/10.1103/PhysRevD.94.104005}}.

 
\bibitem[Barriola and Vilenkin(1989)]{Barriola:1989hx}
Barriola, M.; Vilenkin, A.
\newblock {Gravitational Field of a Global Monopole}.
\newblock {\em Phys. Rev. Lett.} {\bf 1989}, {\em 63},~341.
\newblock {\url{https://doi.org/10.1103/PhysRevLett.63.341}}.

\bibitem[Shi and Li(1991)]{Shi:1991yto}
Shi, X.; Li, X.z.
\newblock {The Gravitational field of a global monopole}.
\newblock {\em Class. Quant. Grav.} {\bf 1991}, {\em 8},~761--767.
\newblock {\url{https://doi.org/10.1088/0264-9381/8/4/019}}.

\end{thebibliography}

\end{document}